\documentclass{bioinfo}
\copyrightyear{2015} \pubyear{2015}

\access{Advance Access Publication Date: Day Month Year}
\appnotes{Manuscript Category}

\usepackage{amsmath}
\usepackage{enumerate}
\usepackage{float}
\usepackage{subcaption}
\usepackage{caption}
\usepackage{dsfont}
\usepackage{graphicx}
\usepackage{natbib}
\usepackage{geometry}
\usepackage{marginnote}
\usepackage{xcolor}
\usepackage{comment}
\usepackage[colorinlistoftodos]{todonotes}

\begin{document}
\firstpage{1}

\subtitle{Subject Section}

\title[Using Data Assimilation to Estimate Glucose and Insulin]{Using Data Assimilation of Mechanistic Models to Estimate Glucose and Insulin Metabolism}
\author[Sample \textit{et~al}.]{Jami J. Mulgrave\,$^{\text{\sfb 1,}}$, Matthew E. Levine $^{\text{\sfb 2}}$, David J. Albers\,$^{\text{\sfb 3}}$, Joon Ha\,$^{\text{\sfb 4}}$, Arthur Sherman\,$^{\text{\sfb 4}}$ and George Hripcsak\,$^{\text{\sfb 1,}*}$}
\address{$^{\text{\sf 1}}$Department of Biomedical Informatics, Columbia University, New York, 10032, USA  \\
$^{\text{\sf 2}}$Department of Computing and Mathematical Sciences, University California Institute of Technology, Pasadena, 91125, USA \\
$^{\text{\sf 3}}$Pediatrics-Informatics and Data Science, University of Colorado Anschutz Medical Campus, Aurora, 80045,
USA and \\
$^{\text{\sf 4}}$Laboratory of Biological Modeling, National Institutes of Health, Bethesda, 20892,
USA}

\corresp{$^\ast$To whom correspondence should be addressed.}

\history{Received on XXXXX; revised on XXXXX; accepted on XXXXX}

\editor{Associate Editor: XXXXXXX}

\abstract{\textbf{Motivation:} There is a growing need to integrate mechanistic models of biological processes with computational methods in healthcare in order to improve prediction. We apply data assimilation in the context of Type 2 diabetes to understand parameters associated with the disease.\\
\textbf{Results:} The data assimilation method captures how well patients improve glucose tolerance after their surgery.  Data assimilation has the potential to improve phenotyping in Type 2 diabetes.\\
\textbf{Contact: } \href{hripcsak@columbia.edu}{hripcsak@columbia.edu}\\
}

\maketitle

\section{Introduction}
There is a growing need to integrate mechanistic models of biological processes with computational methods in healthcare.  Popular computational methods, such as machine learning, are useful for predicting outcomes of interest, yet are limited in settings with sparse, irregular, and inaccurate data.  Adding mechanistic models to machine learning methods aims to boost the power of the analyses by adding physiological constraints and minimizing the data required.  

Data assimilation \citep{law_data_2015} combines mechanistic models with data using Bayesian statistics to make forecasts.  Data assimilation (DA) has been successfully employed in diverse fields including the geosciences \citep{carrassi_data_2018} and biomedicine \citep{tang_data_2018}. We build upon previous work \citep{albers_personalized_2017} and \citep{albers_mechanistic_2018} that applies data assimilation in the context of Type 2 diabetes. \cite{albers_personalized_2017} discussed how  data assimilation could be used to forecast future glucose values, to impute previously missing glucose values, and to infer Type 2 diabetes phenotypes. \cite{albers_mechanistic_2018} showed that data assimilation forecasts compare well with specific glucose measurements and match or exceed in accuracy expert forecasts.  These findings could be of potential use for diabetes self-management.  We are interested in using data assimilation to improve phenotyping in Type 2 diabetes.

We use oral glucose tolerance tests (OGTTs) to demonstrate how data assimilation could be used to estimate the parameters of a model of glucose and insulin dynamics.  OGTTs are commonly used to diagnose diabetes and can also identify patients with impaired glucose tolerance.  OGTT data can be found in an electronic medical record and can be compared between patients and studied over time.  We would like to study the differences in parameter patterns between patients with normal glucose, glucose intolerance, and diabetes using the lab results of OGTTs.  Better understanding of a patient’s underlying physiological parameters could in principle lead to better understanding of diabetes and eventually better treatments.





\section{Materials}

We created a dataset of OGTTs for each patient using the electronic medical record at Columbia University.   We first pulled together all of the laboratory tests related to glucose measurements that could be relevant to OGTTs.  We selected only male patients to avoid measuring parameters related to gestational diabetes.  We selected all male patients that had at least two dates of glucose measurements and at least three glucose measurements per date.  We kept all glucose measurements that appeared to be related to OGTTs.  We also pulled together all of the laboratory tests related to insulin measurements that occurred on the same dates as the glucose measurements. This initial dataset resulted in 200 male patients out of 564750 male patients with glucose data.  

Out of the 200 patients, we kept glucose measurements that occurred at the regularly expected times of an OGTT, which are combinations of fasting, 30 minute, 1 hour, 1.5 hour, 2 hour, 3 hour, 4 hour, 5 hour, and 6 hour.  The most frequent combinations were fasting, 30 minute, 2 hour, and fasting, 60 minute, and 2 hour.  We assumed that at these times, the patients received 75 grams of  glucose after taking a fasting glucose measurement.  We kept the insulin measurements that occurred at the timepoints of the glucose measurements.  We removed patients that had glucose measurements that were not consistent with OGTTs, had repeated measurements, or were missing glucose measurements.  We kept patients that had missing insulin measurements as long as they had complete glucose measurements. We kept patients that had at least two OGTTs.  As a result of these conditions, we had a dataset of 147 patients that we used for the analysis. We collected any hemoglobin A1C (HbA1c) data that were available for these patients.

\begin{methods}
\section{Methods}
Data assimilation (DA) is a method that combines models with data to reconstruct
the model state and provide forecasts.  We use a longitudinal mathematical model described in \cite{ha_type_2019} that is capable of representing the metabolic state of an individual at any point in time during their progression from normal glucose tolerance to Type 2 diabetes over a period of years.  We aim to reconstruct the parameters relevant to glucose-insulin
dynamics from the oral glucose tolerance test measurements of patients using the mathematical model. 


We solve seven ordinary differential equations (ODE) in the data assimilator.  We use the built-in MATLAB solver, \textit{ode45}, to solve these equations.  These ODEs are displayed in Equations \ref{eq1} -- \ref{eq7}. Please review \citet{ha_type_2019} for more details on the equations.

\begin{equation}
\label{eq1}
    \frac{dG}{dt} = MEAL + HGP  - (E_{GO} + S_II)G 
\end{equation}

\begin{equation}
\label{eq2}
    \frac{dI}{dt} = \frac{\beta}{V}ISR - kI 
\end{equation}
    
\begin{equation}
\label{eq3}
    ISR = \sigma \frac{(M + \gamma)^{kISR}}{\alpha_{ISR}^{kISR}(M + \gamma)^{kISR}} 
\end{equation}

\begin{equation}
\label{eq4}
    M = \frac{G^{kM}}{\alpha_M^{kM} + G^{kM}} 
\end{equation}

\begin{equation}
\label{eq5}
    \frac{d\gamma}{dt} = \frac{\gamma_{\infty}(G)\gamma}{\tau_{\gamma}}  
\end{equation}

\begin{equation}
\label{eq6}
    \frac{d\sigma}{dt} = \frac{\sigma_{\infty}(ISR,M) - \sigma}{\tau_{\sigma}}  
\end{equation}

\begin{equation}
\label{eq7}
    \frac{d\beta}{dt} = \frac{(P(ISR)- A(M))\beta)}{\tau_{\beta}}
\end{equation}

Equation \ref{eq1} is the glucose $(G)$ equation.  It says that $G$ increases as a result of meal influx ($MEAL$) and hepatic glucose production $(HGP)$ and decreases as a result of uptake. The parameter $I$ represents insulin and $S_I$ represents insulin sensitivity.  The parameter $E_{GO} $ represents disposal.

Equation \ref{eq2} is the insulin $I$ equation.  It says that $I$ decreases due to removal with rate constant $k$ and increases due to secretion by beta cells, where $\beta$ is the beta-cell mass, described in Equation \ref{eq7}.  $ISR$ is the insulin secretion rate described in Equation \ref{eq3} and $V$ is the volume of distribution. 

Equation \ref{eq3} describes further the insulin secretion rate ($ISR$).  The value $M$ represents beta-cell metabolism and is described further in Equation \ref{eq4}.  The parameter $\gamma$ represents the effect of K(ATP) channel density to shift the glucose dependence of secretion.  The parameter $\sigma$ represents insulin secretion, described further in Equation \ref{eq6}.  

Equation \ref{eq4} describes beta-cell metabolism, $M$, where $M$ is assumed to be a sigmoidally-increasing function of $G$. 

Equation \ref{eq5} describes the parameter $\gamma$, where $\gamma_{\infty}$ is an increasing sigmoidal function of $G$, and $\tau_{\gamma}$ is the time constant.  

Equation \ref{eq6} describes beta-cell functional compensation where it is assumed that increased $ISR$ leads to an increase in $\sigma$ whereas increased $M$ leads to a decrease in $\sigma$.

Equation \ref{eq7} describes increased beta-cell mass, $\beta$.  It is assumed $\beta$ is increased by proliferation, $P$, and decreased by apoptosis, $A$.  

We estimate two parameters, $\sigma$, (sigma) and $S_I $ (SI), combining deterministic optimization using interior-point methods with stochastic optimization using Monte Carlo Markov chain (MCMC). Specifically we use deterministic methods to quickly survey the solution surface to set both initial values and prior boundaries for the sigma and SI parameters for the MCMC. The final estimation of parameters and the uncertainty quantification of those estimates is calculated using the standard formulation of random walk Metropolis Hastings algorithm \citep{metropolis_equation_1953}. Each patient is estimated using three MCMC chains run with 10000 iterations.  We choose the parameter estimates using the chain that minimized the mean squared error.  The MCMC proposal step size coefficient was 0.1 for all parameters. The sigma and SI parameters had good convergence of the chains as seen in Figure \ref{fig:00}. 

We apply upper and lower boundaries to estimate the parameters.  We use $0<\textup{sigma}<5000$ and  $0 < \textup{SI} < 5$. All of the other parameters in the ODE equations were set to nominal values, values that have been reported in the literature.  We consider the effect of insulin on the results by running the data assimilation with and without insulin.

\begin{figure}[!tpb]
\centerline{\includegraphics[width = 0.5\textwidth]{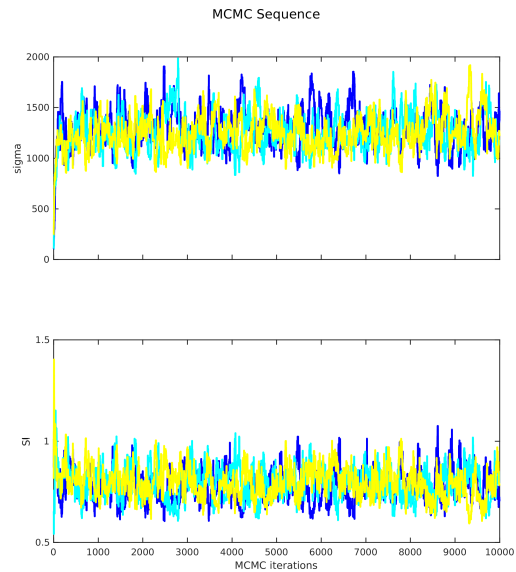}}
\caption{Example of a Diagnostic Plot of Three Chains for Sigma and SI}\label{fig:00}
\end{figure}

\enlargethispage{6pt}

\end{methods}

\section{Results}

We reviewed the medical histories of nine of the 147 patients that were run in the data assimilation.  We were interested to learn if their histories coincided with the parameters that were being estimated in the model. We used the guidelines of the National Institute of Diabetes and Digestive and Kidney Diseases \citep{american_diabetes_association_classification_2016} and \citep{bergman_petition_2018} to determine whether the OGTT was normal, impaired glucose (prediabetes) or diabetes.  Since we collected HbA1c data, we also reviewed the disease states as determined by the HbA1c value, which is also detailed in \citep{american_diabetes_association_classification_2016}.  We show in Tables \ref{Tab:01} -- \ref{Tab:02} the test numbers that are needed for each diagnosis.

\begin{table}[!t]
\begin{tabular}{@{}llll@{}}\toprule Diagnosis &
Fasting & 1 Hour Glucose & 2 Hour Glucose\\\midrule
Normal & 99 or below & 154 or below & 139 or below\\
Impaired Glucose & 100 to 125 & 155 and above & 140 to 199 \\
Diabetes & 126 or above &  & 200 or above\\\botrule
\end{tabular}
\caption{Diagnosing Diabetes using the Oral Glucose Tolerance Test}
\label{Tab:01}
\end{table}

\begin{table}[!t]
\begin{tabular}{@{}ll@{}}\toprule Diagnosis &
 A1C Level \\\midrule
Normal & below 5.7 percent \\
Impaired Glucose & 5.7 to 6.4 percent \\
Diabetes & 6.5 percent or above\\\botrule
\end{tabular}
\caption{Diagnosing Diabetes using the A1C Test}
\label{Tab:02}
\end{table}




For the analysis, we calculated average values for the sigma, SI, and sigma * SI parameters. We used a burnin of 5000 for each of the three chains and averaged the remaining iterations across the chains for each parameter.

We found that overall, for these nine patients, there was clear separation between the parameters, sigma, SI, and sigma*SI, and disease, when including and excluding insulin.  In addition, the values for the parameters are generally higher for normal glucose and lower for impaired and diabetic glucose.

In particular, we consider the results of the nine patients when including insulin.  For the sigma parameter, there were four patients that had separation and for those patients, the values for the sigma parameter was higher for the normal glucose than for the impaired and diabetic glucose.  Two patients did not have clear separation of the parameter and three  patients did not have at least two disease levels to tell if there was separation.  For the SI parameter, there were five patients that had separation of the parameter.  For three of these cases, the values for the SI parameter for the normal glucose was lower than the values for the impaired and diabetic glucose.  In the other two cases, the values for the SI parameter for the normal glucose was higher than the impaired and diabetic glucose. There was one patient that did not have clear separation and three patients did not have at least two disease levels to tell if there was separation.  For the sigma * SI parameter, there were six patients with clear separation of the parameters.  For those patients, the values for the sigma*SI parameter was higher for the normal glucose than for the impaired and diabetic glucose.  For the other three patients, they did not have at least two disease levels to tell if there was separation.

We consider the results of the nine patients when not including insulin.  For the sigma parameter, there were four patients that separation, and for those patients, two patients had values for the sigma parameter that was higher for the normal glucose than the impaired and diabetic glucose, and two patients had values for the sigma parameter that was lower for the normal glucose than the impaired and diabetic glucose.  Three patients did not have clear separation of the parameter and two patients did not have at least two disease levels to tell if there was separation.  For the SI parameter, there were seven patients that had separation.  For all seven patients, the values for the SI parameter were higher for the normal glucose than for the impaired and diabetic glucose.  Two patients did not have at least two disease levels to tell if there was separation.  For the sigma * SI parameter, there were six patients that had separation.  For those patients, the values for the sigma * SI parameter were higher for the normal glucose than for the impaired and diabetic glucose.  One patient did not have clear separation of the parameter and two patients did not have at least two disease levels to tell if there was separation.

\begin{figure}[!tpb]
\centerline{\includegraphics[width = 0.6\textwidth]{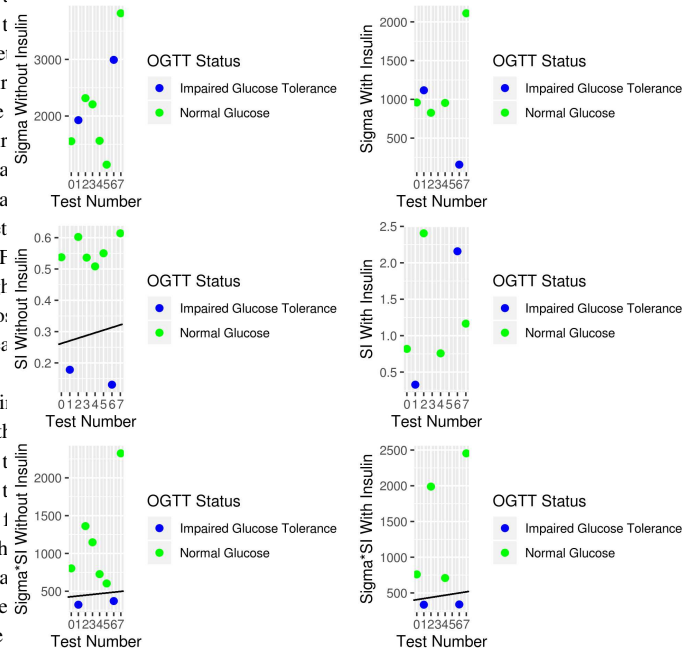}}
\caption{Scatterplots of Parameters by Test Number}\label{fig:01}
\end{figure}

\begin{figure}[!tpb]
\centerline{\includegraphics[width = 1\linewidth]{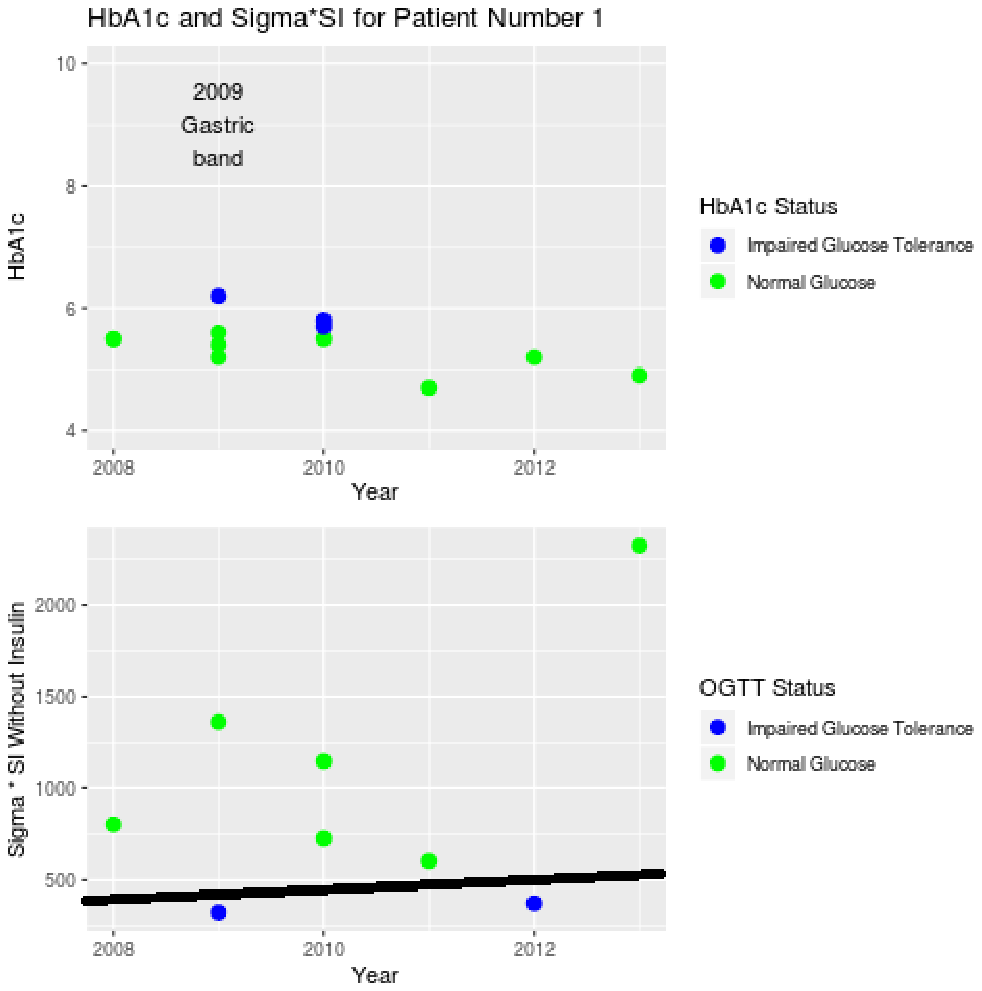}}
\caption{Scatterplots of HbA1c and Sigma * SI  by Year}\label{fig:02}
\end{figure}

\begin{figure}[!tpb]
\centerline{\includegraphics[width = 1\linewidth]{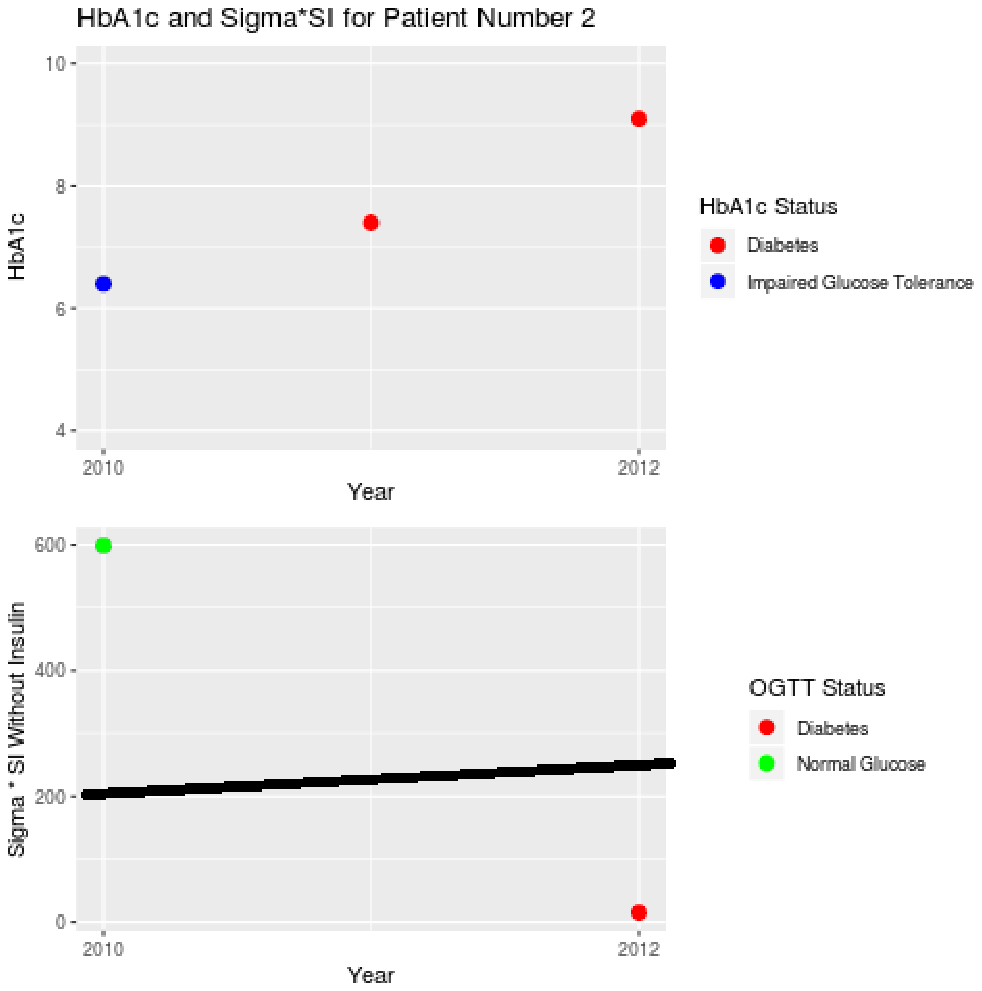}}
\caption{Scatterplots of HbA1c and Sigma * SI  by Year}\label{fig:03}
\end{figure}

\begin{figure}[!tpb]
\centerline{\includegraphics[width = 1\linewidth]{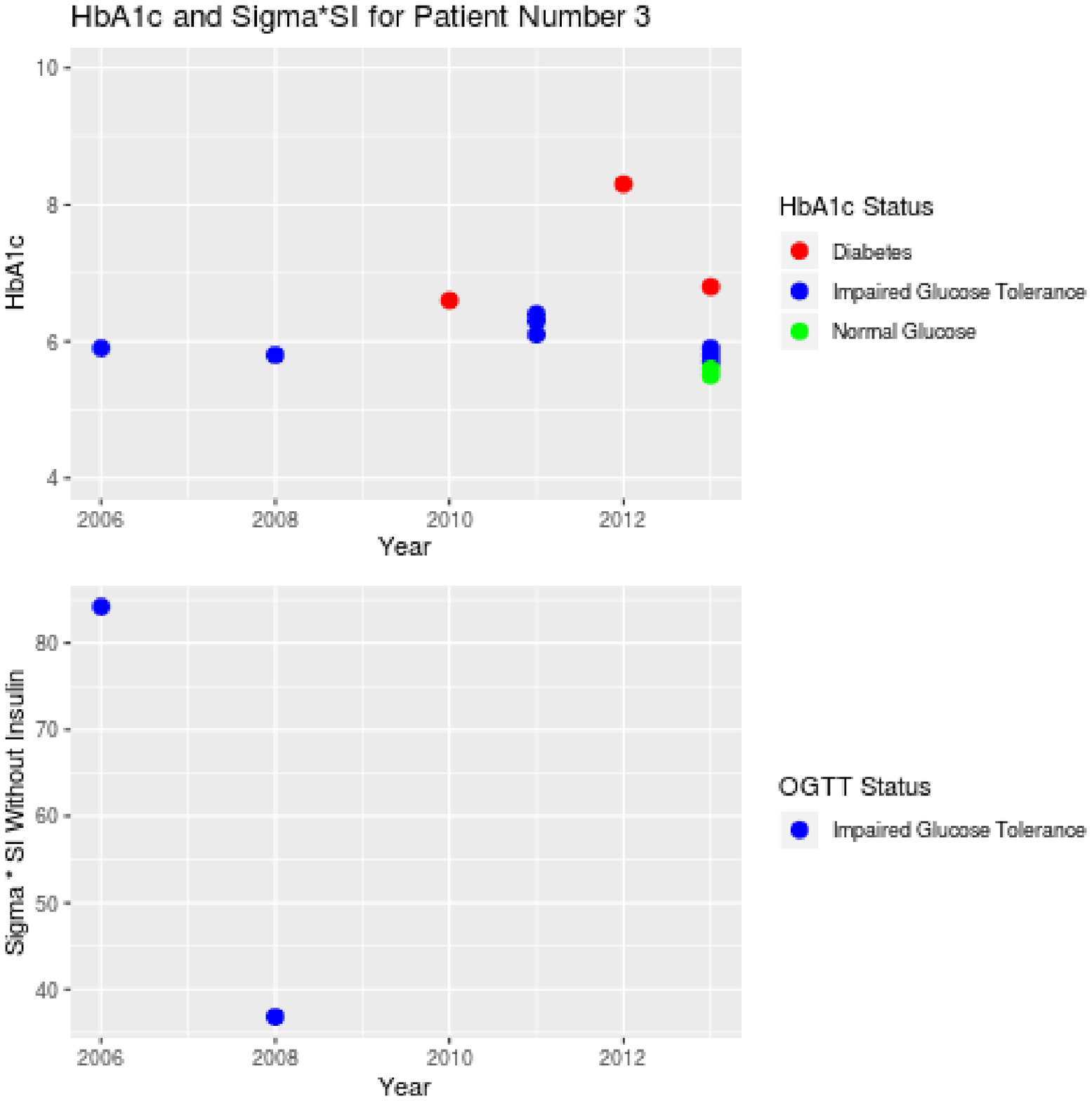}}
\caption{Scatterplots of HbA1c and Sigma * SI by Year}\label{fig:04}
\end{figure}

\begin{figure}[!tpb]
\centerline{\includegraphics[width = 1\linewidth]{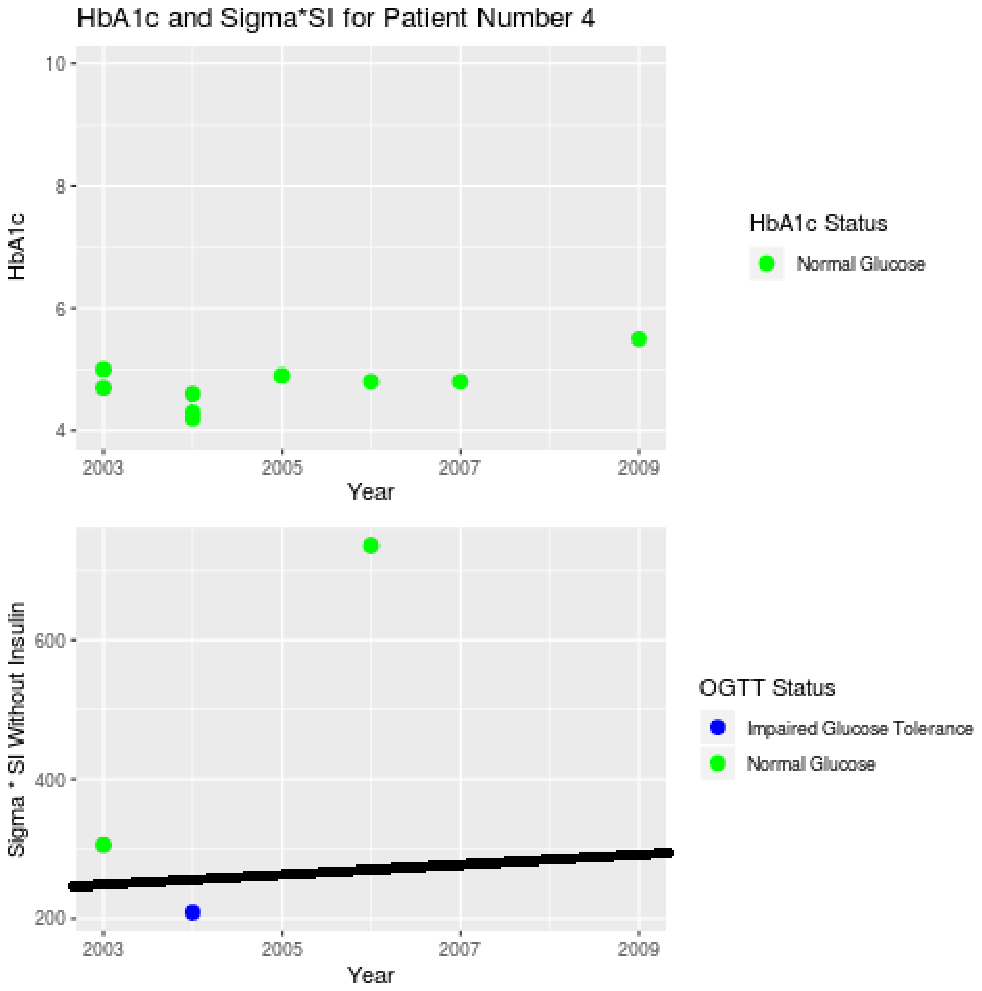}}
\caption{Scatterplots of HbA1c and Sigma * SI by Year}\label{fig:05}
\end{figure}

\begin{figure}[!tpb]
\centerline{\includegraphics[width = 1\linewidth]{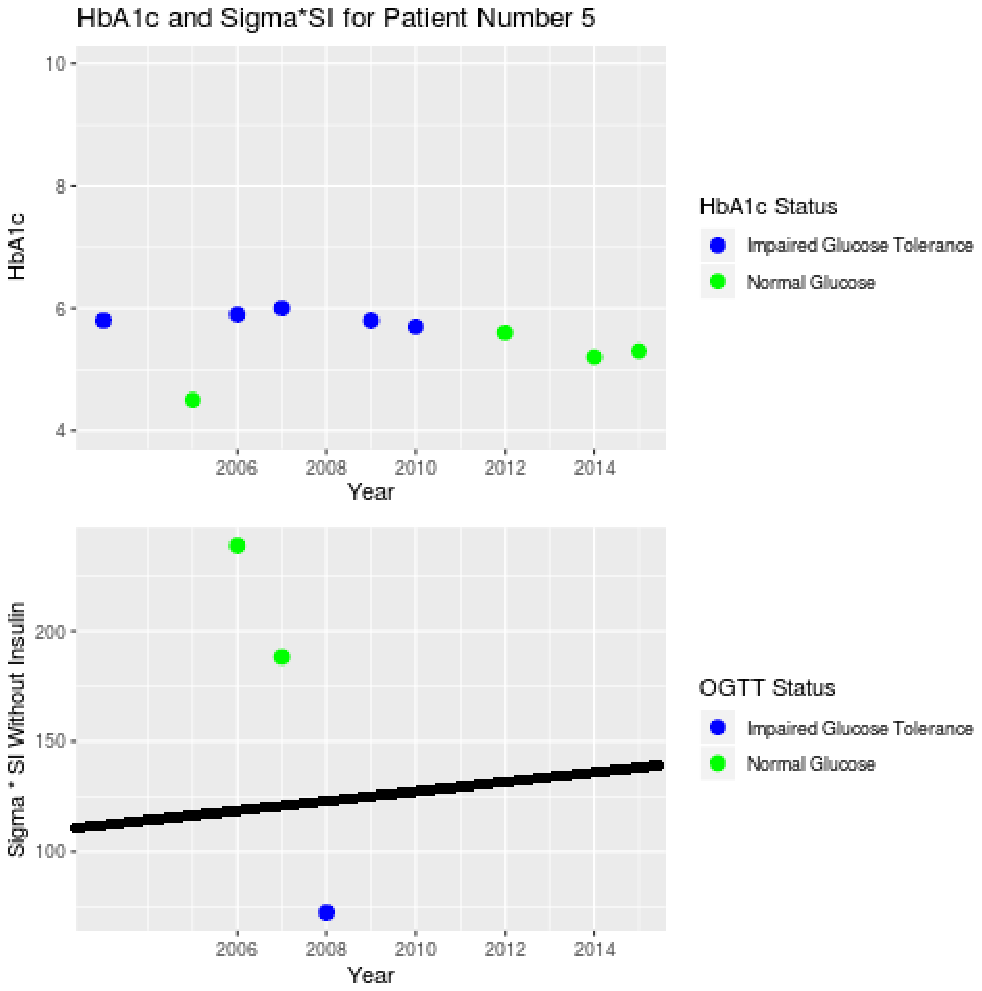}}
\caption{Scatterplot of Parameters and HbA1c by Year}\label{fig:06}
\end{figure}

\begin{figure}[!tpb]
\centerline{\includegraphics[width = 1\linewidth]{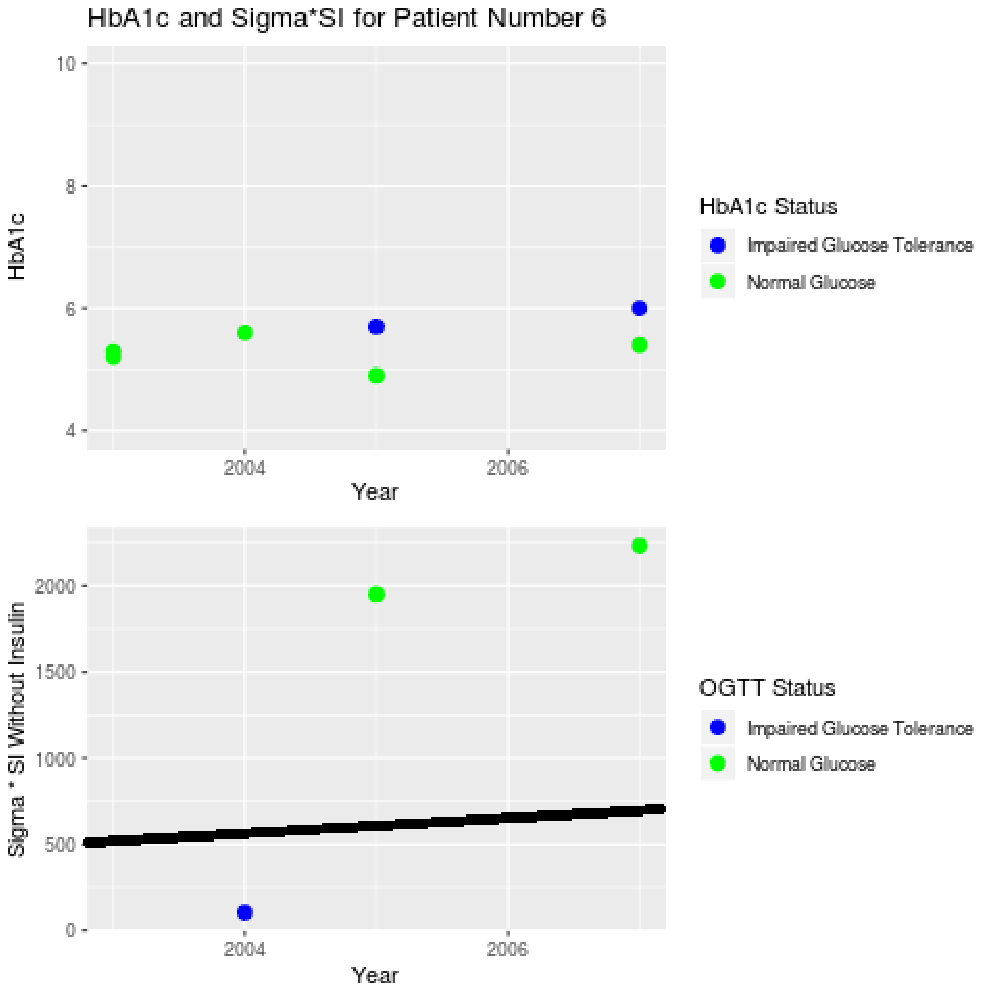}}
\caption{Scatterplot of Parameters and HbA1c by Year}\label{fig:07}
\end{figure}

\begin{figure}[!tpb]
\centerline{\includegraphics[width = 1\linewidth]{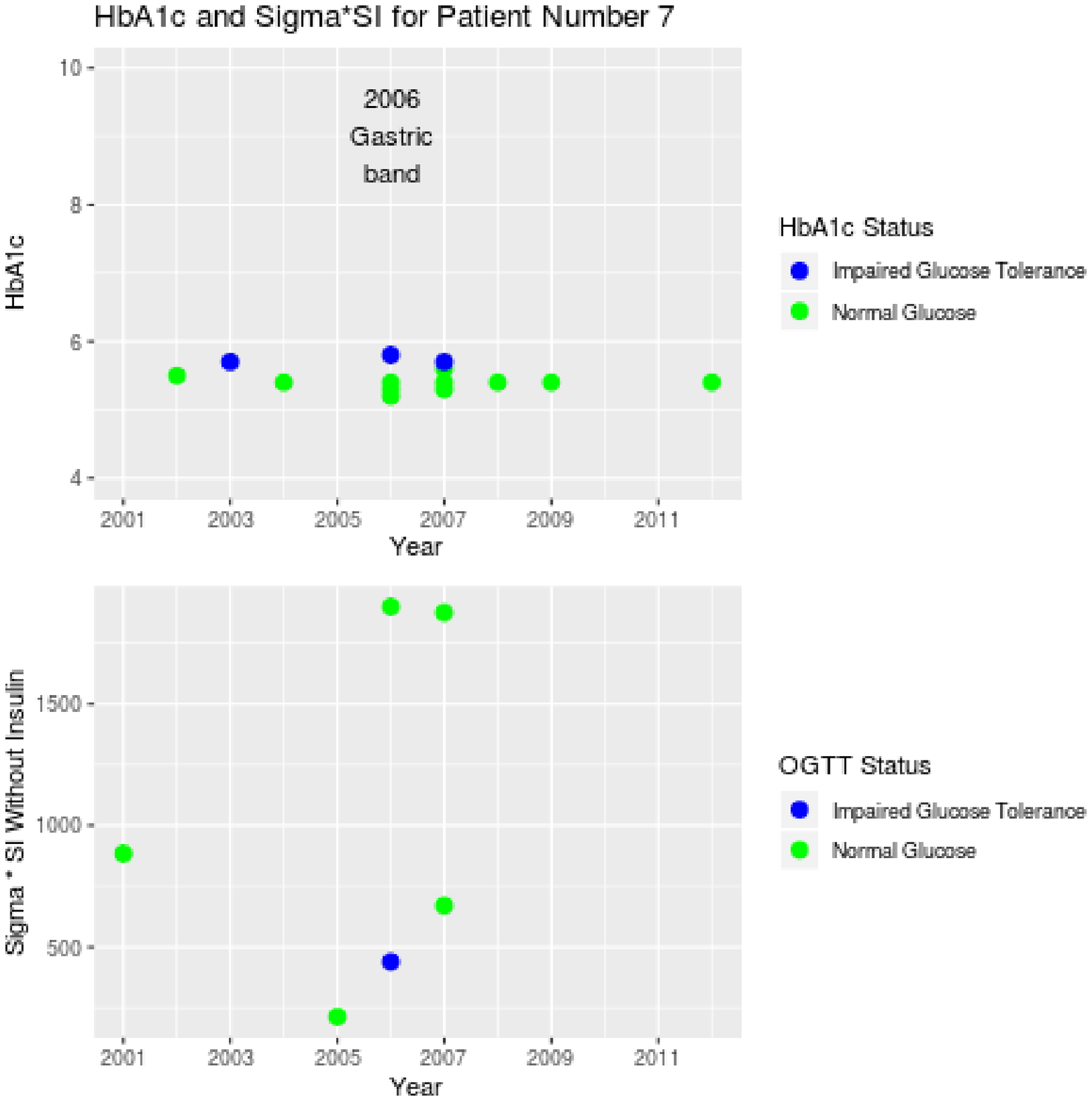}}
\caption{Scatterplots of HbA1c and Sigma * SI by Year}\label{fig:08}
\end{figure}

\begin{figure}[!tpb]
\centerline{\includegraphics[width = 1\linewidth]{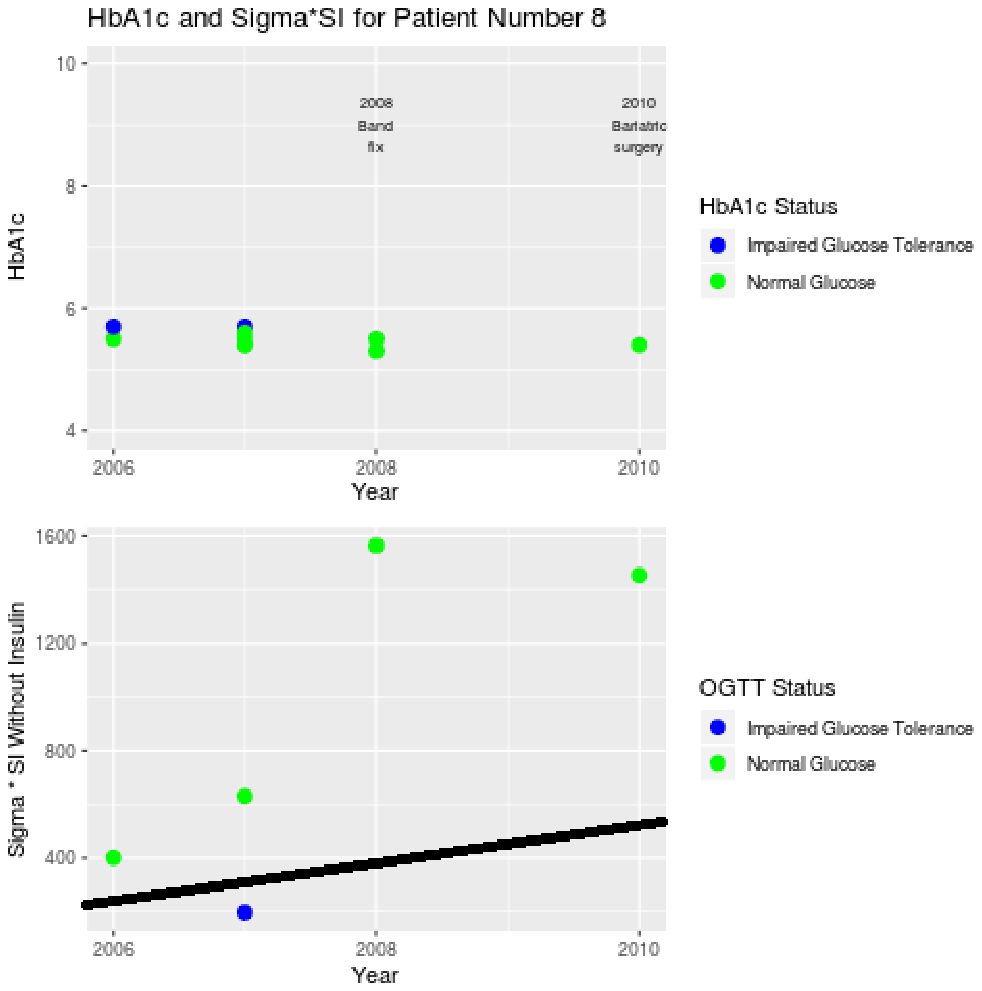}}
\caption{Scatterplot of Parameters and HbA1c by Year}\label{fig:09}
\end{figure}

\begin{figure}[!tpb]
\centerline{\includegraphics[width = 1\linewidth]{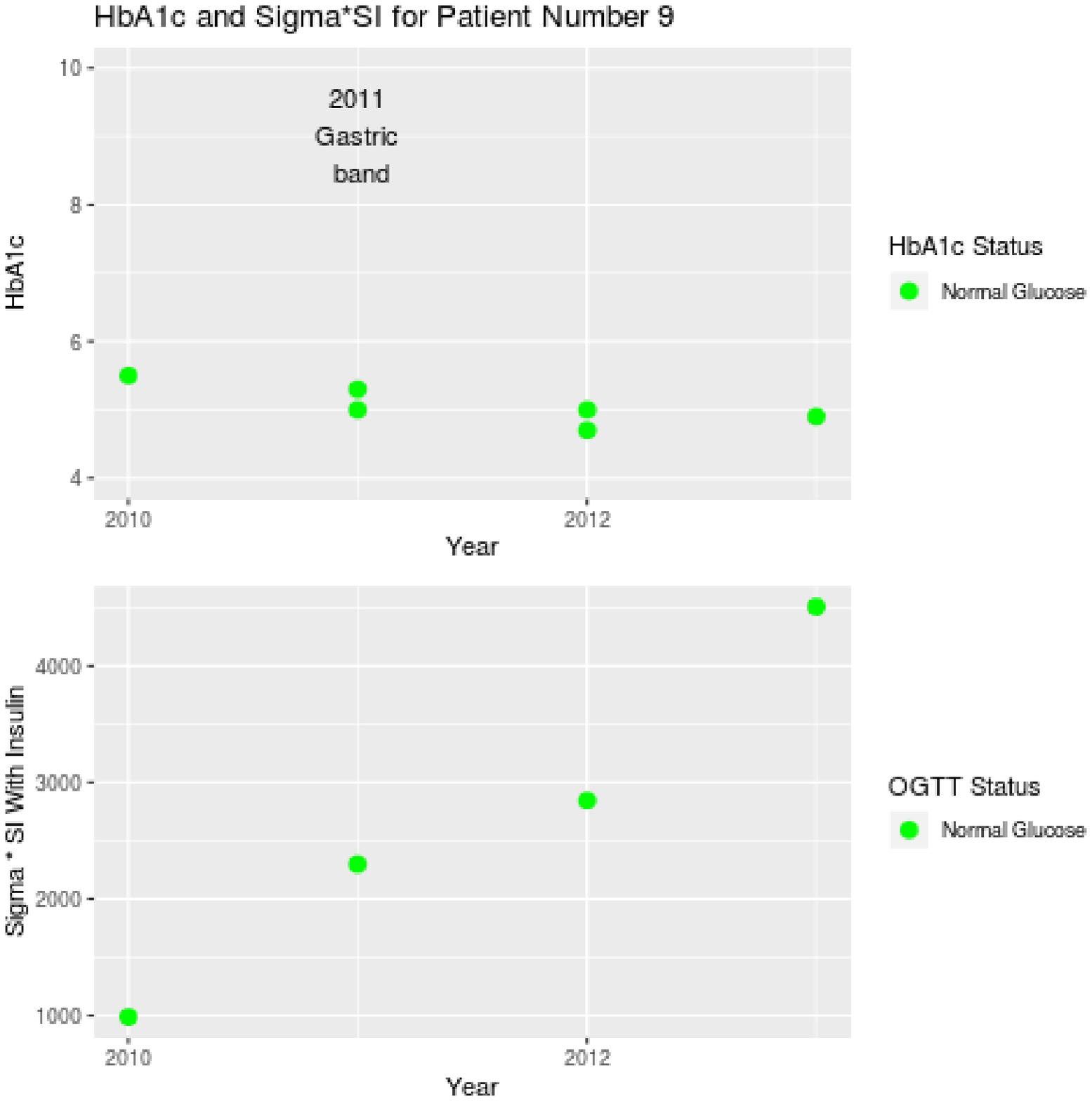}}
\caption{Scatterplots of HbA1c and Sigma * SI by Year}\label{fig:10}
\end{figure}


Figure \ref{fig:01} is a sample patient, Patient 1, from the nine patients, that displays the sigma, SI, and sigma * SI parameter values, with and without insulin, for each OGTT.  This patient had eight OGTTs, ordered from Test Number 0 to Test Number 7.  The sigma parameter does not have clear separation, with or without insulin, and the SI parameter with insulin does not have clear separation.  The SI parameter without insulin and the sigma * SI parameter, with and without insulin, has clear separation of values between normal and impaired glucose.  Figures \ref{fig:02} -- \ref{fig:10} show sigma * SI parameter values, without insulin, for the nine patients of which we reviewed their medical histories.  

For Figure \ref{fig:02}, for Patient 1, there is clear separation of the sigma * SI parameter values, without insulin, between normal and impaired glucose. The sigma * SI values are higher for the normal glucose than for the impaired glucose.  Patient 1 had a gastric band in 2009, and at the next OGTT which occurred in 2009, the sigma * SI improved.  The improvement is indicated by the higher green value for 2009 in the Figure \ref{fig:02}.  The sigma * SI also improved after the second impaired glucose tolerance result which occurred 2012.  The HbA1c values for Patient 1 were either normal or impaired glucose.

For Figure \ref{fig:03}, for Patient Number 2, there is clear separation of the sigma * SI parameters,  without insulin.    The sigma * SI values are higher for the normal glucose than for the diabetic glucose.  The HbA1cs were either impaired or diabetic glucose.  

For Figure \ref{fig:04}, for Patient Number 3, there is no clear separation because the OGTTs results only showed impaired glucose tolerance.  Also, this patient did not have insulin measurements.  The HbA1cs were normal, impaired, and diabetic glucose.

For Figure \ref{fig:05}, for Patient Number 4, there clear separation of sigma * SI, without insulin.  The sigma * SI values are higher for the normal glucose than for the impaired glucose.  The HbA1cs are all normal  glucose.  

For Figure \ref{fig:06}, for Patient Number 5, there clear separation of sigma * SI, without insulin.  The sigma * SI values are higher for the normal glucose than for the impaired glucose.  The HbA1c values were either normal or impaired glucose.

For Figure \ref{fig:07}, for Patient Number 6, there clear separation of sigma * SI, without insulin.  The sigma * SI values are higher for the normal glucose than for the impaired glucose.  The HbA1c values were either normal or impaired glucose.

For Figure \ref{fig:08}, for Patient Number 7, we are unable to tell if there is separation of the sigma * SI parameters, without insulin, because there is a sigma * SI value for the normal glucose that is lower than the impaired glucose.  This patient had an impaired glucose tolerance result and subsequently, a gastric band in 2006.  After the gastric band, the patient had a normal glucose result in 2006 and an improved sigma * SI value, as evident by the higher green value in the graph for the year 2006.  The HbA1c values were normal or impaired glucose. 

For Figure \ref{fig:09}, for Patient Number 8, there clear separation of sigma * SI, without insulin.  The sigma * SI values are higher for the normal glucose than for the impaired glucose.  This patient had a band fix in 2008 and bariatric surgery in 2010.  After the band fix, the sigma * SI improved and in the year of the bariatric surgery, the sigma * SI remained high. The HA1c values are normal or impaired glucose.  

Lastly, for Figure \ref{fig:10}, for Patient Number 9, we are unable to tell if there is separation of the sigma * SI parameter since all of the OGTTs were normal glucose.  This patient had a gastric band in 2011 and afterwards, the sigma * SI parameter improved every OGTT afterwards.  

In the Appendix, we placed the Figures \ref{fig:App01} -- \ref{fig:App09} for the with insulin case of the nine patients. These figures show the sigma * SI parameter values and the HbA1c values in the year that they occurred.  The separation results are similar to the without insulin cases.   

Thus, considering gastric related surgeries, such as the gastric band, gastric band fix, and bariatric surgery, the sigma * SI improves after one to two OGTTs.  In addition, there appears to be clear separation of sigma * SI for normal, impaired glucose, and diabetic glucose, with and without insulin.  Generally, the sigma * SI values are higher for the normal glucose than for the impaired and diabetic glucose.

\begin{figure}[!tpb]
\centerline{\includegraphics[width = 1\linewidth]{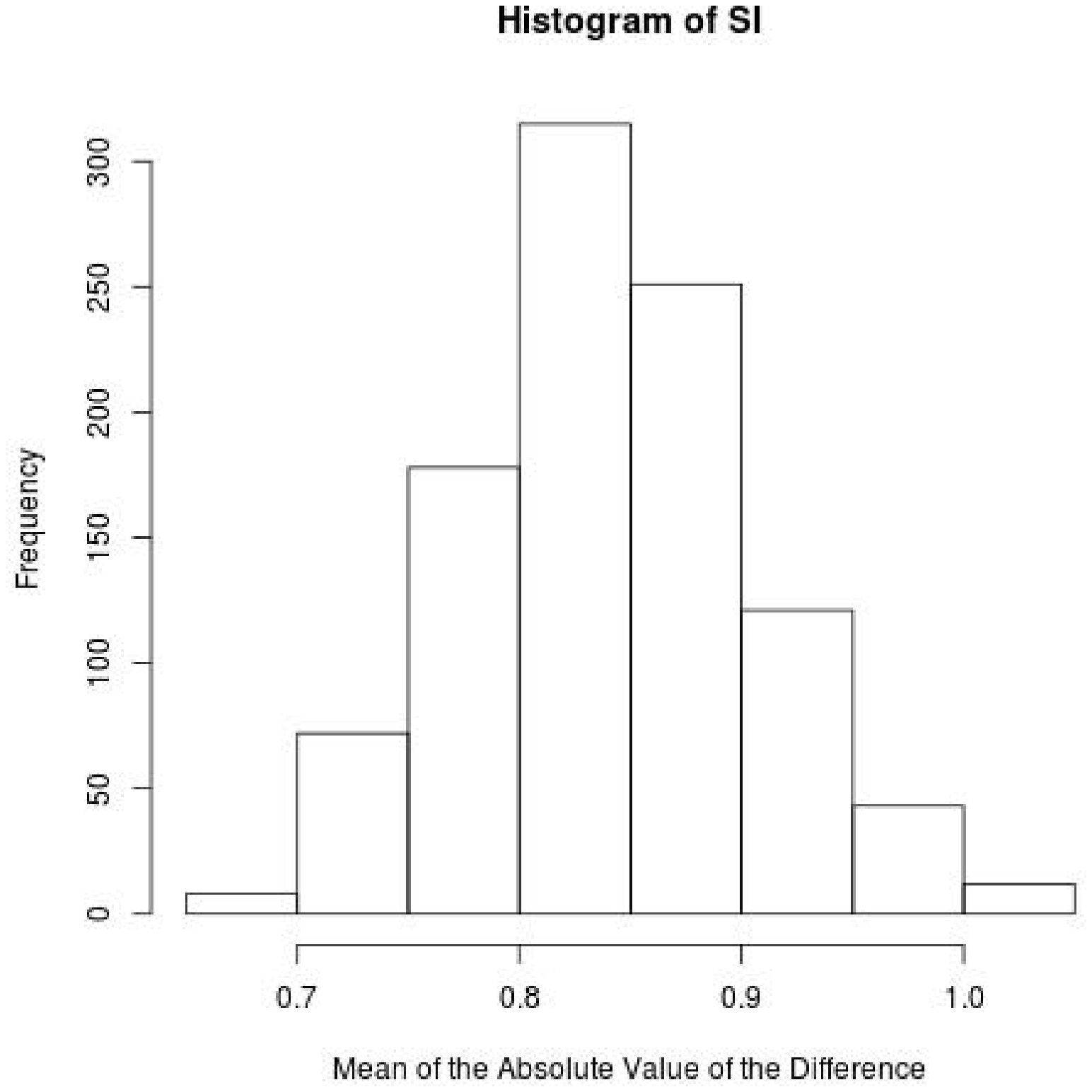}}
\caption{Histogram of the Mean of the Absolute Value of the Differences of SI}\label{fig:11}
\end{figure}

\begin{figure}[!tpb]
\centerline{\includegraphics[width = 1\linewidth]{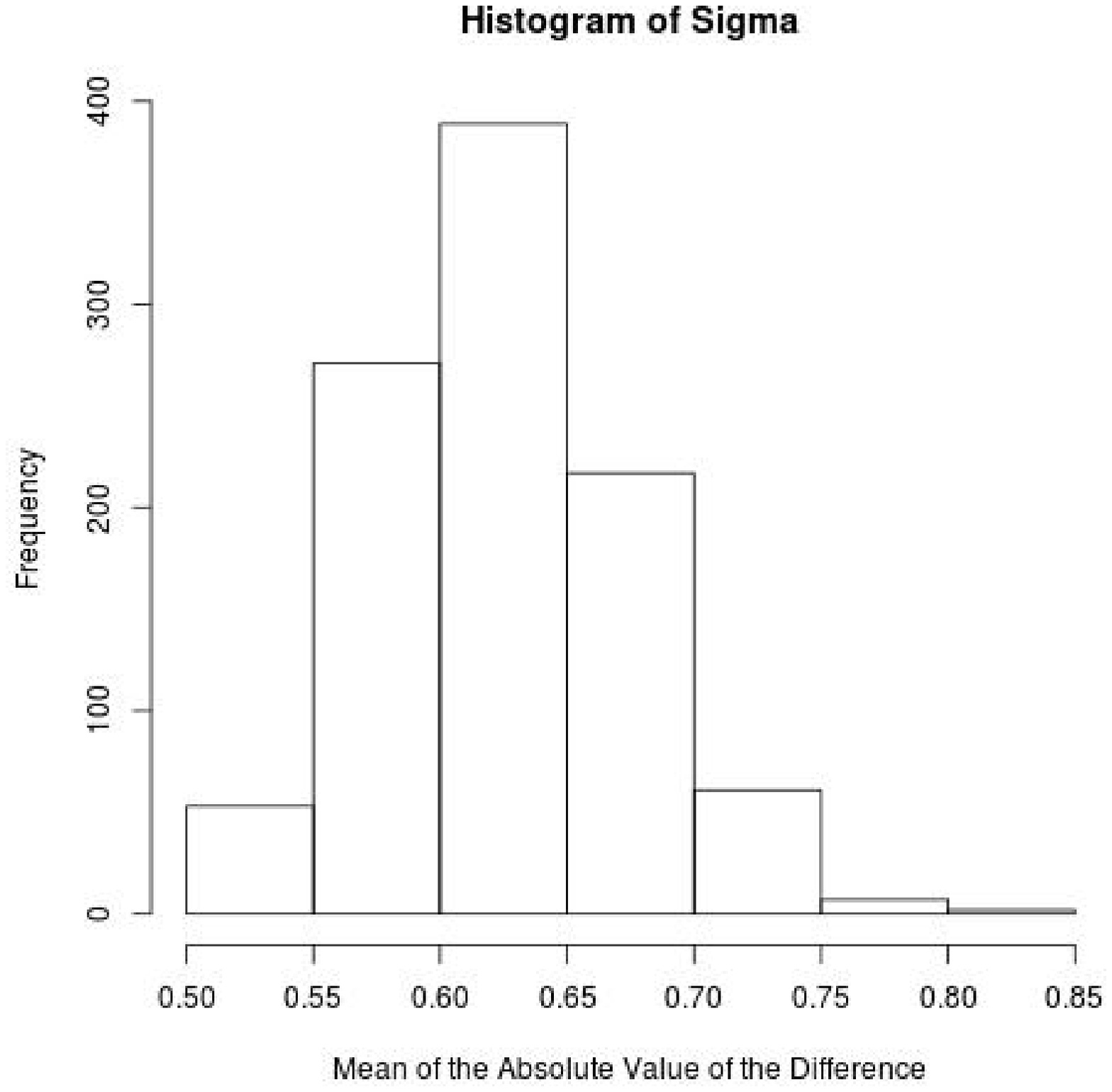}}
\caption{Histogram of the Mean of the Absolute Value of the Differences of SI}\label{fig:12}
\end{figure}

\begin{figure}[!tpb]
\centerline{\includegraphics[width = 1\linewidth]{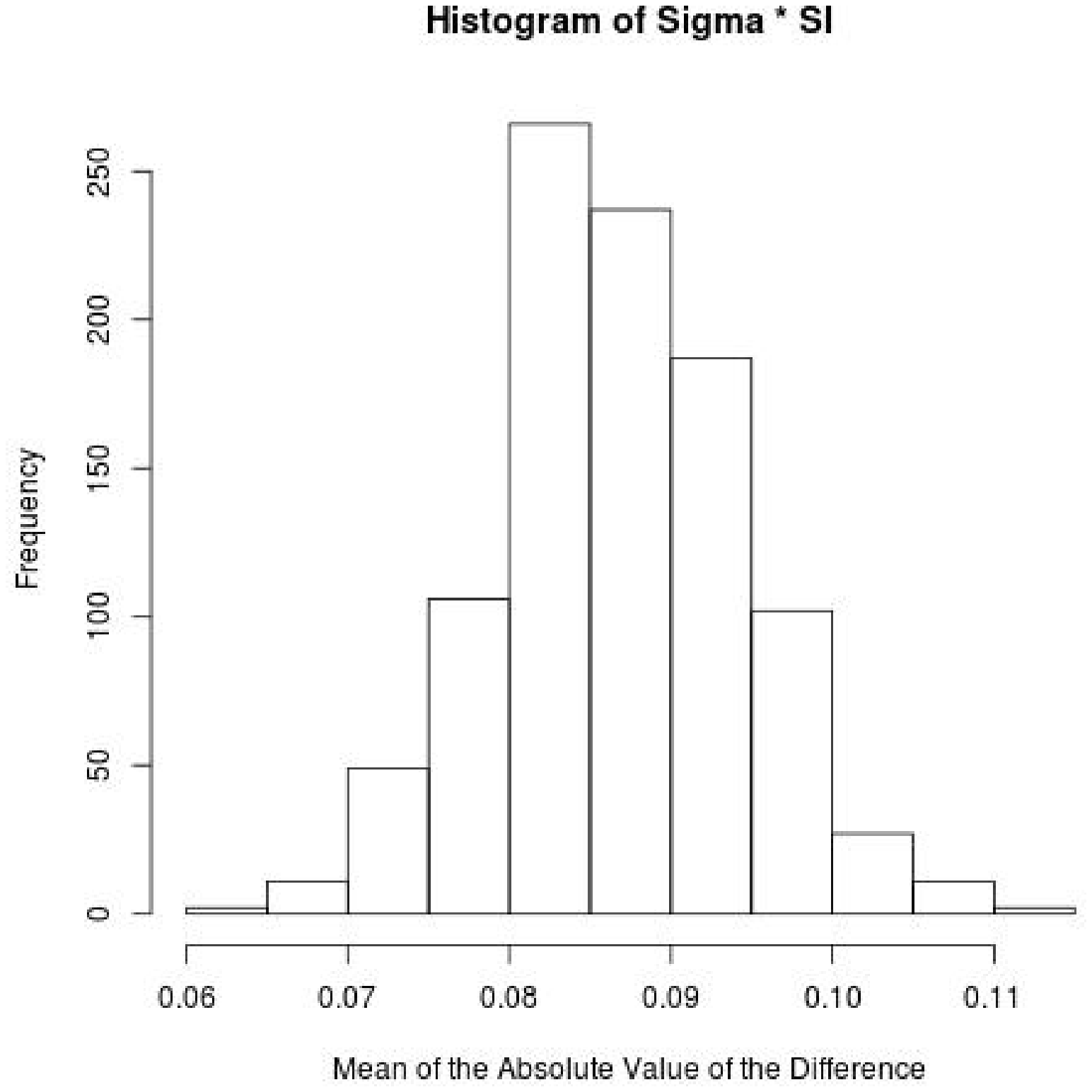}}
\caption{Histogram of the Mean of the Absolute Value of the Differences of SI}\label{fig:13}
\end{figure}

We used a quantitative analysis to review the results for all of the patients.  Since there appear to be differences in the parameters given the inclusion or exclusion of insulin, we wish to compare the change in sigma, the change in SI, and the change in sigma * SI and determine the parameter that is the most robust to the change.  There were 124 patients with OGTTs that had both glucose and insulin, so we used those patients for the comparison.  We standardized the dataset by scaling and centering the parameters in order to compare the parameters to each other meaningfully.  Since we are interested in differences, we create a new variable of interest, the absolute value of the difference, for sigma, SI, and sigma * SI.   Specifically, we calculated the absolute value of the difference of sigma with and without insulin, of SI with and without insulin, and of sigma * SI with and without insulin.  We used a clustered bootstrap to draw the samples for each of these variables of interest.  We clustered the bootstrap since each patient could have multiple OGTTs.  To cluster, we sampled with replacement the Study ID numbers and included all measurements associated with those IDs.  The statistic of interest was the mean of the absolute value of the differences.  We collected 1000 sets of bootstrap samples of size 124 of the absolute value of the differences for each parameter and we found the mean of each of the 1000 sets.  Thus, we have three clustered bootstrap distributions of the mean of the absolute value of the differences.  Using each bootstrap distribution, we can calculate the mean, standard error, and 95\% confidence interval and compare the results.  Please see Table \ref{Tab:03} for the results.  Please see Figures \ref{fig:11} -- \ref{fig:13} for the distributions of the mean of the absolute value of the differences.

\begin{table}[!t]
\begin{tabular}{@{}llll@{}}\toprule Parameter &
Mean & Standard Error & 95\% Confidence Interval\\\midrule
Sigma & 0.6254 & 0.0015 & [0.6224, 0.6284]\\
SI &  0.8411 & 0.0020 &  [0.8371, 0.8451]\\
Sigma * SI & 0.0868 & 0.0002 & [0.0864, 0.0872]\\\botrule
\end{tabular}
\caption{Statistics for the Bootstrap Distributions}
\label{Tab:03}
\end{table}

Based on the results, all parameters are significantly affected by the exclusion of insulin.  Sigma * SI has the smallest change in values when including and excluding insulin, compared to sigma, and SI.  SI has the largest change when including and excluding insulin.  Thus, sigma * SI is more robust to the exclusion of insulin than sigma and SI.  SI is most affected by the exclusion of insulin.  

\begin{figure}[!tpb]
\centerline{\includegraphics[width = 1\linewidth]{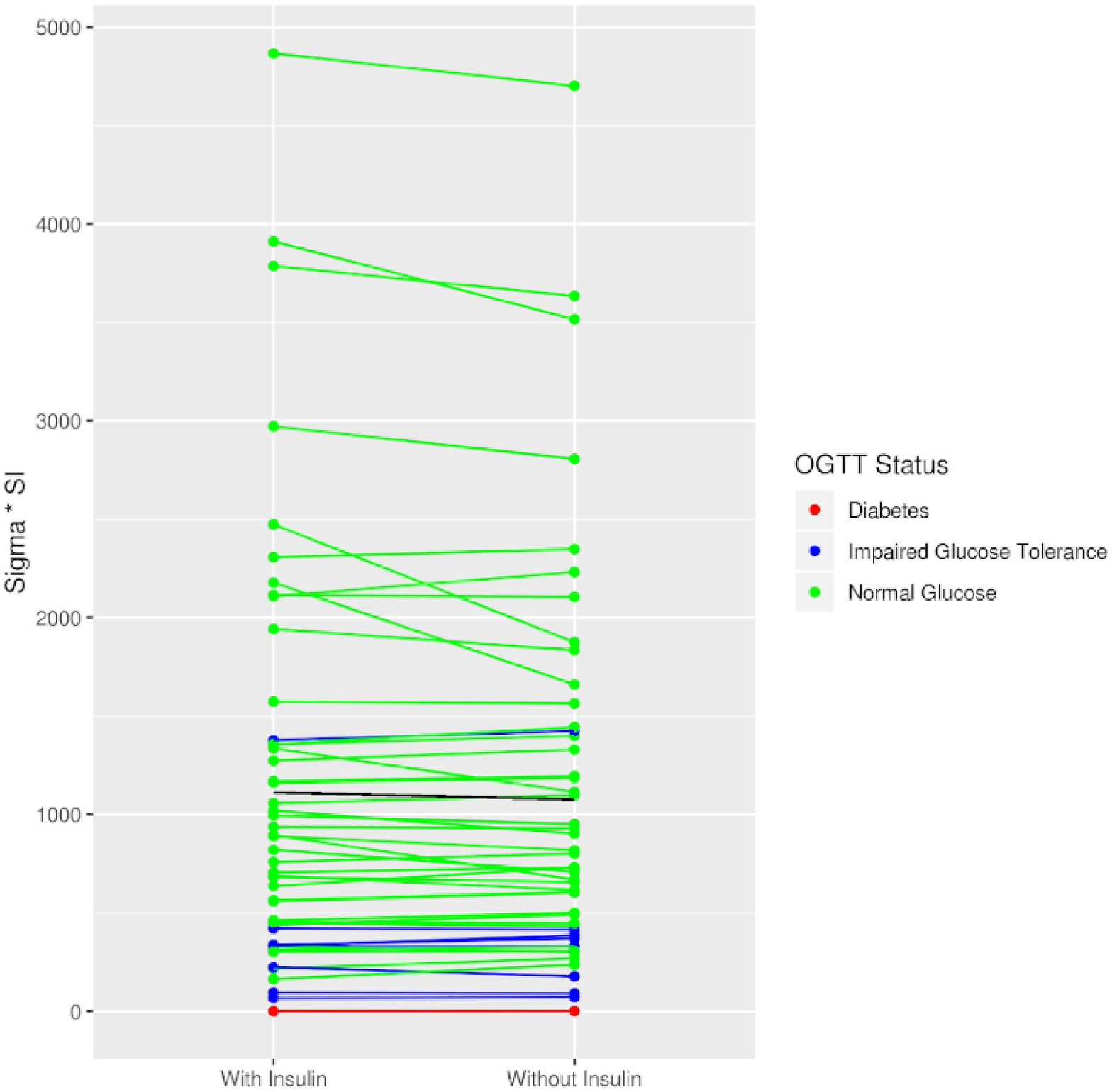}}
\caption{Change of Sigma * SI}\label{fig:14}
\end{figure}

\begin{figure}[!tpb]
\centerline{\includegraphics[width = 1\linewidth]{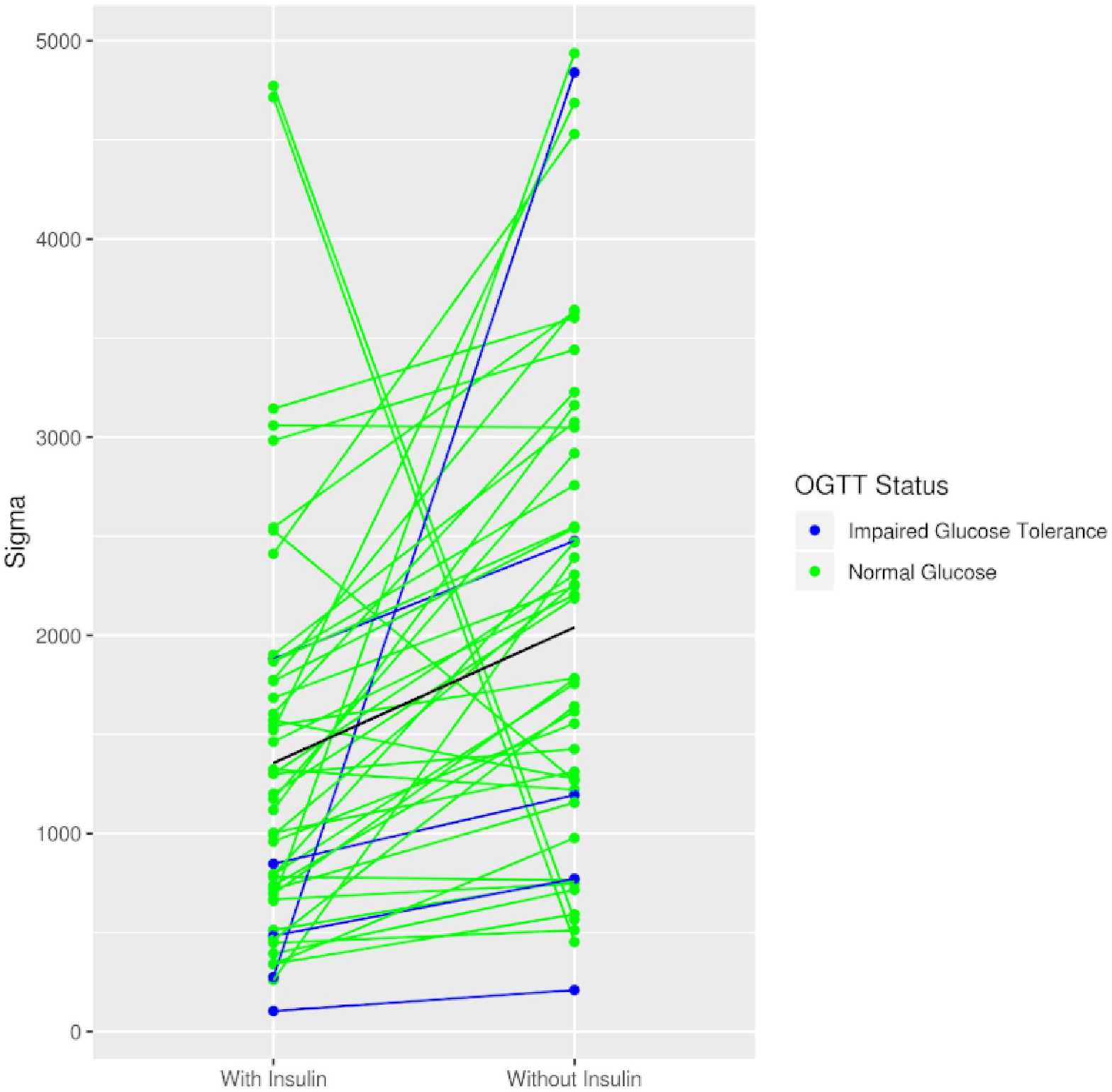}}
\caption{Change of Sigma}\label{fig:15}
\end{figure}

\begin{figure}[!tpb]
\centerline{\includegraphics[width = 1\linewidth]{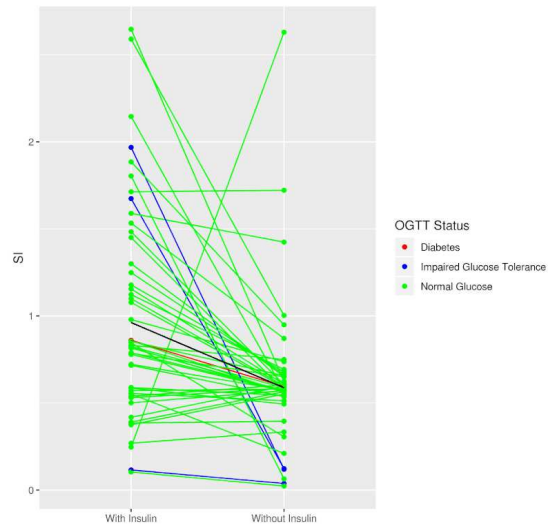}}
\caption{Change of SI}\label{fig:16}
\end{figure}

Lastly, we wished to understand the direction of the relationship between the parameters when including insulin.  In particular, we looked at the direction of the change in sigma * SI, the direction of the change in sigma, and the direction of the change in SI, when including and excluding insulin.  Using the original data, we calculated the means of each parameter, with and without insulin.  These means are displayed in Table \ref{Tab:04}.  We plotted a random sample of 50 OGTTs from the original data set.  We also plotted, in black, a line connecting the mean value for all of the OGTTs for each parameter with insulin and without insulin.  These plots are displayed in Figures \ref{fig:14} -- \ref{fig:16}. In Figure \ref{fig:14}, we show the plot for sigma * SI.  On average, there little change in sigma * SI when including and excluding insulin, as displayed by the flat slope of the line.  In contrast, there are much larger changes in sigma and changes in SI.  In Figure \ref{fig:15}, there is a large positive change in sigma when removing insulin.  In Figure \ref{fig:16}, there is a large negative change in SI when removing insulin.  The steepness of the change for both sigma and SI are similar.

\begin{table}[!t]
\begin{tabular}{@{}lll@{}}\toprule Parameter &
Mean With Insulin & Mean Without Insulin \\\midrule
Sigma &  1355.136 &2039.622\\
SI &   0.962 &  0.588 \\
Sigma * SI & 1111.803 & 1076.999 \\\botrule
\end{tabular}
\caption{Group Means for Each Parameter}
\label{Tab:04}
\end{table}

\section{Discussion}
Data assimilation is a promising method that could be used in healthcare to improve predictions.  We applied data assimilation in the context of Type 2 diabetes with the hope of using the method to improve phenotyping.  We have learned from the results that the data assimilation method captures how well patients improve glucose tolerance after their surgery.  After a patient has a surgery, there is a big jump in sigma * SI.  In this way, the data assimilation method captures information that doctors would not ordinarily see.   In addition, the sigma * SI values are more robust to the exclusion of insulin than sigma and SI alone.  Thus, even if insulin measurements are not collected, which commonly occurs when patients have an oral glucose tolerance test, doctors can still learn information about a patient's diabetic disease. 

In the future, it would be interesting to analyze the patterns of improvement in glucose tolerance as body mass index (BMI) changes after surgery.  It would also be interesting to see if we could predict the measurements of the next OGTT.  Lastly, it would be interesting to run the data assimilation method in the cases of missing glucose and insulin measurements.

%
%



\section*{Funding}

This work was funded in part by National Institutes of Health grants R01 LM006910 and T15 LM007079.  The work was also funded in part by National Library of Medicine R01 LM012734.

\bibliographystyle{apalike}
\bibliography{Zotero.bib}
%
%









\newpage
\appendix

\begin{figure}[!tpb]
\centerline{\includegraphics[width = 1\linewidth]{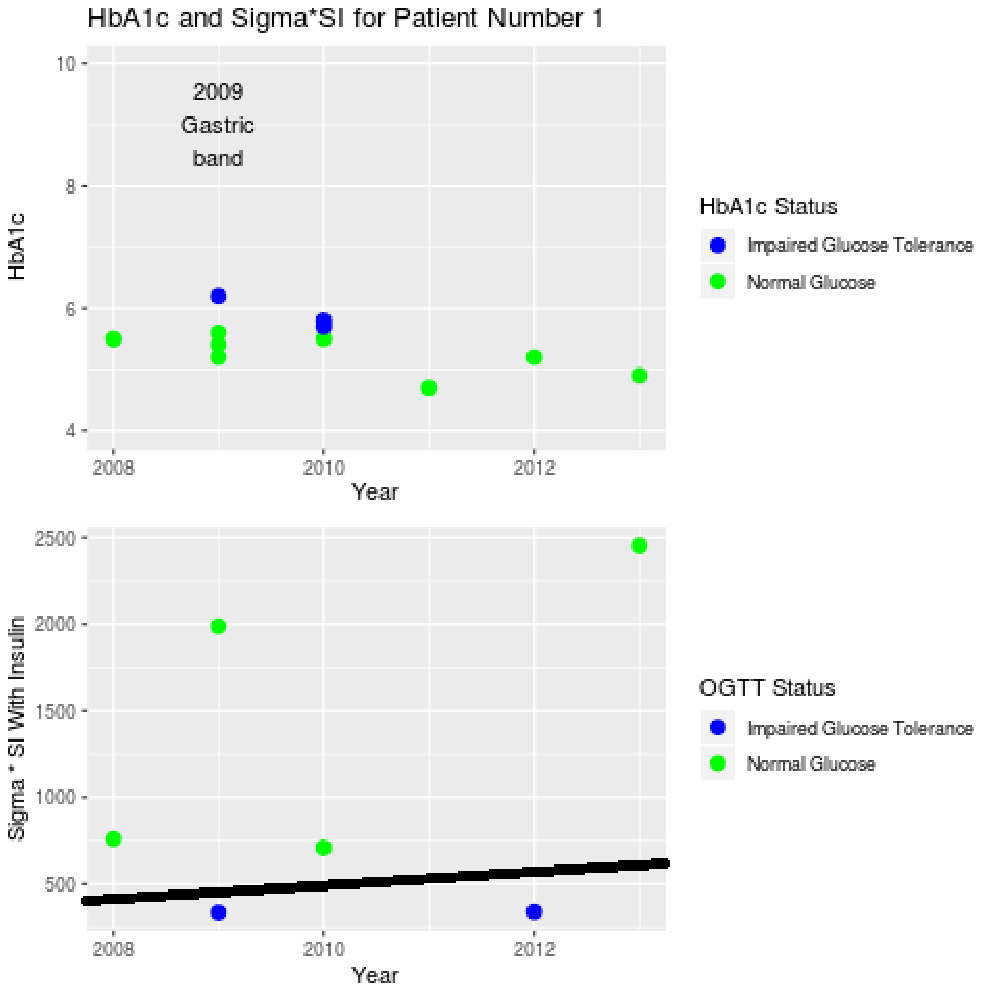}}
\caption{Scatterplot of Parameters and HbA1c by Year}\label{fig:App01}
\end{figure}

\begin{figure}[!tpb]
\centerline{\includegraphics[width = 1\linewidth]{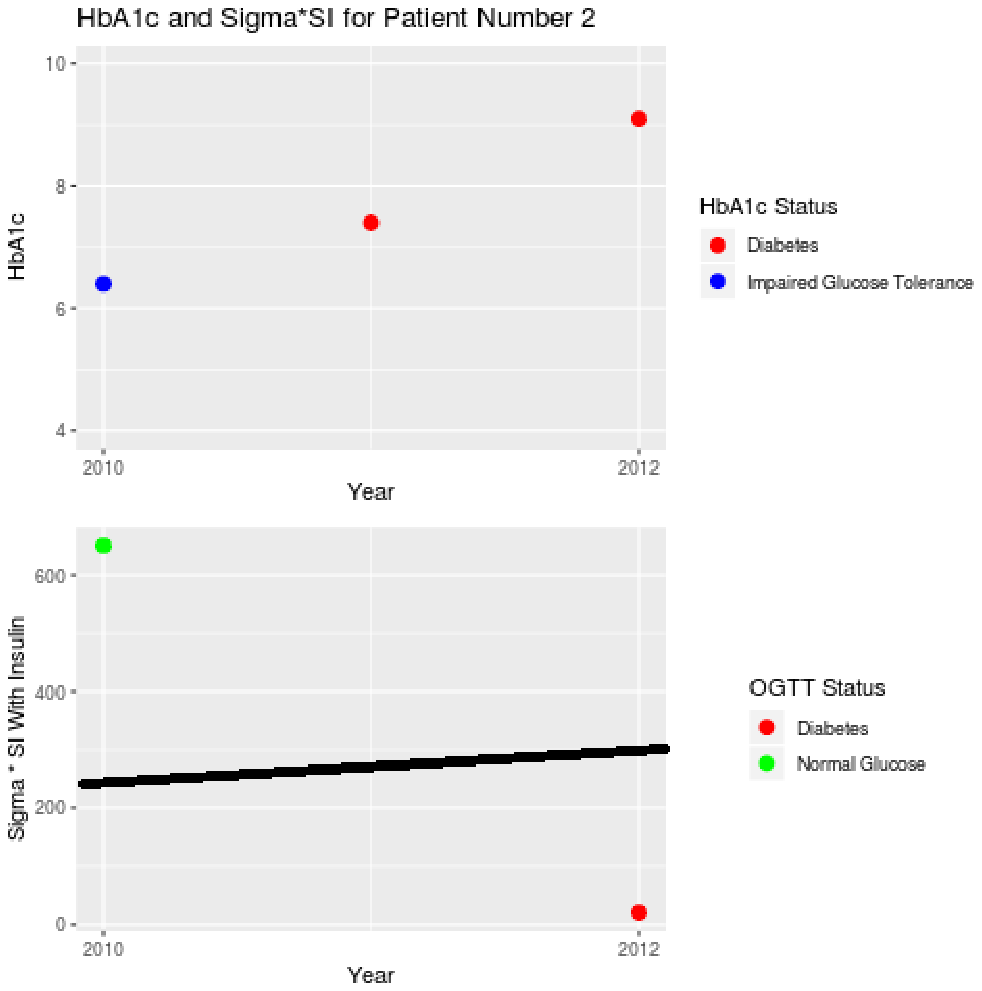}}
\caption{Scatterplot of Parameters and HbA1c by Year}\label{fig:App02}
\end{figure}

\begin{figure}[!tpb]
\centerline{\includegraphics[width = 1\linewidth]{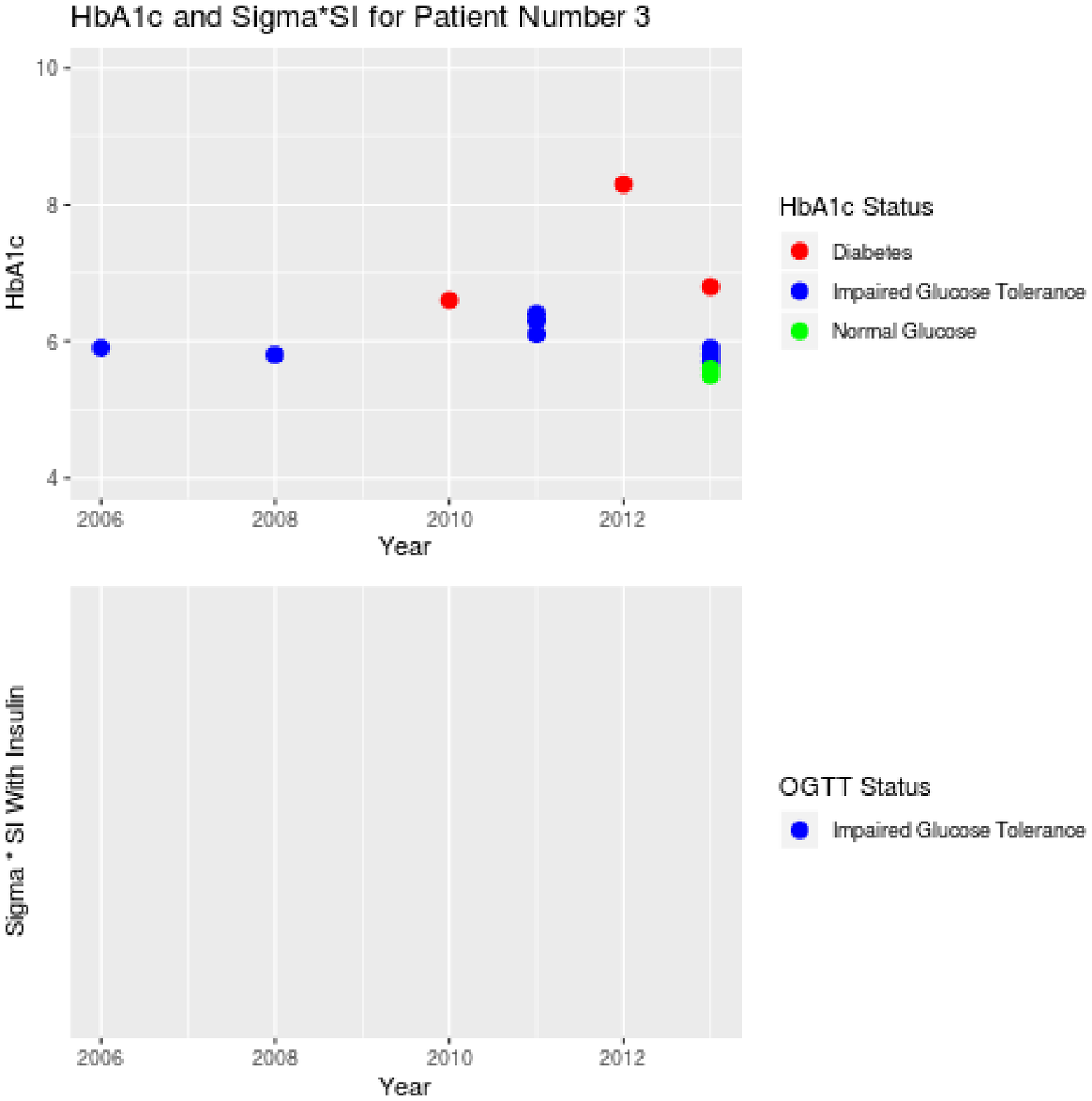}}
\caption{Scatterplot of Parameters and HbA1c by Year}\label{fig:App03}
\end{figure}

\begin{figure}[!tpb]
\centerline{\includegraphics[width = 1\linewidth]{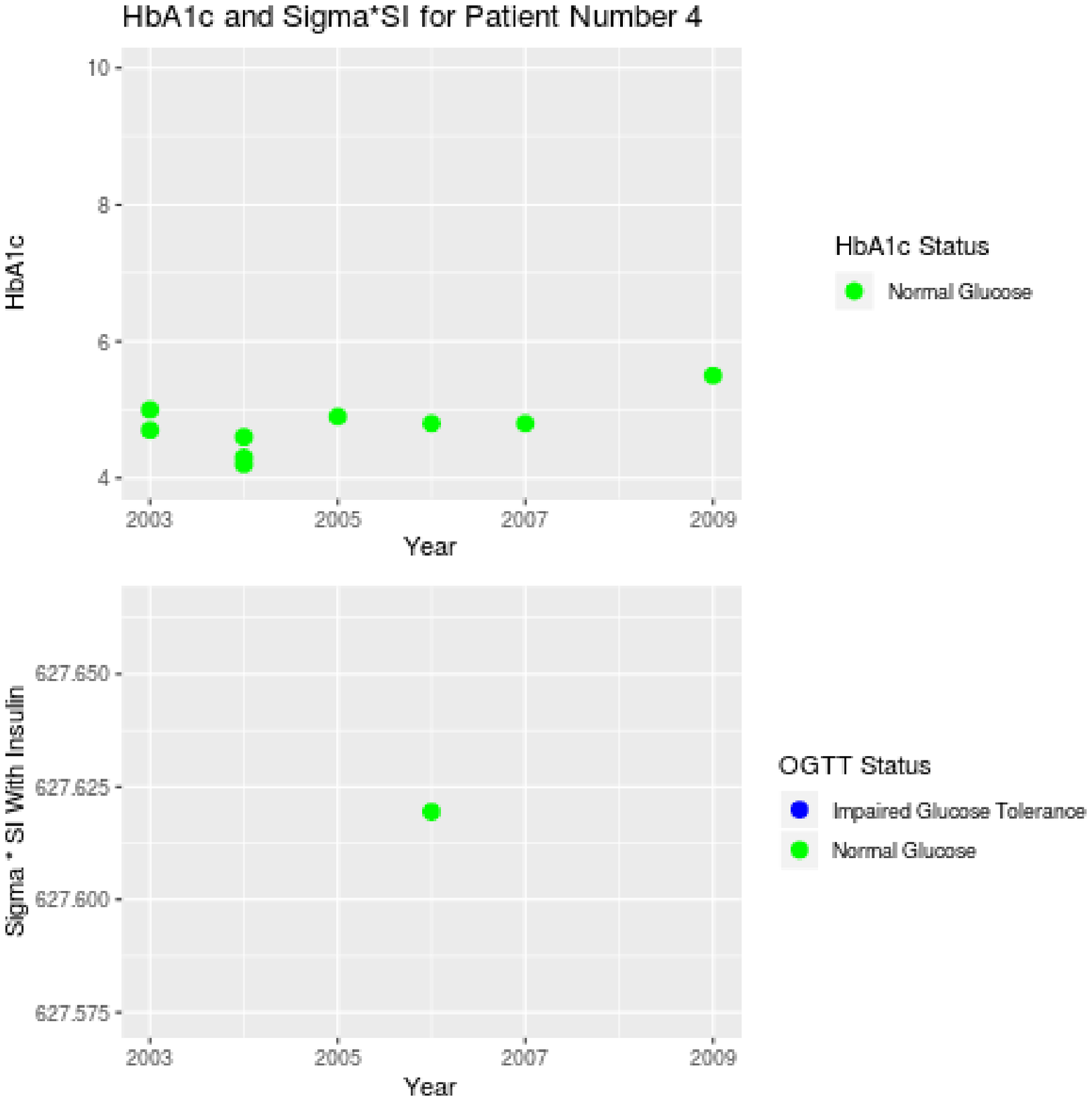}}
\caption{Scatterplot of Parameters and HbA1c by Year}\label{fig:App04}
\end{figure}

\begin{figure}[!tpb]
\centerline{\includegraphics[width = 1\linewidth]{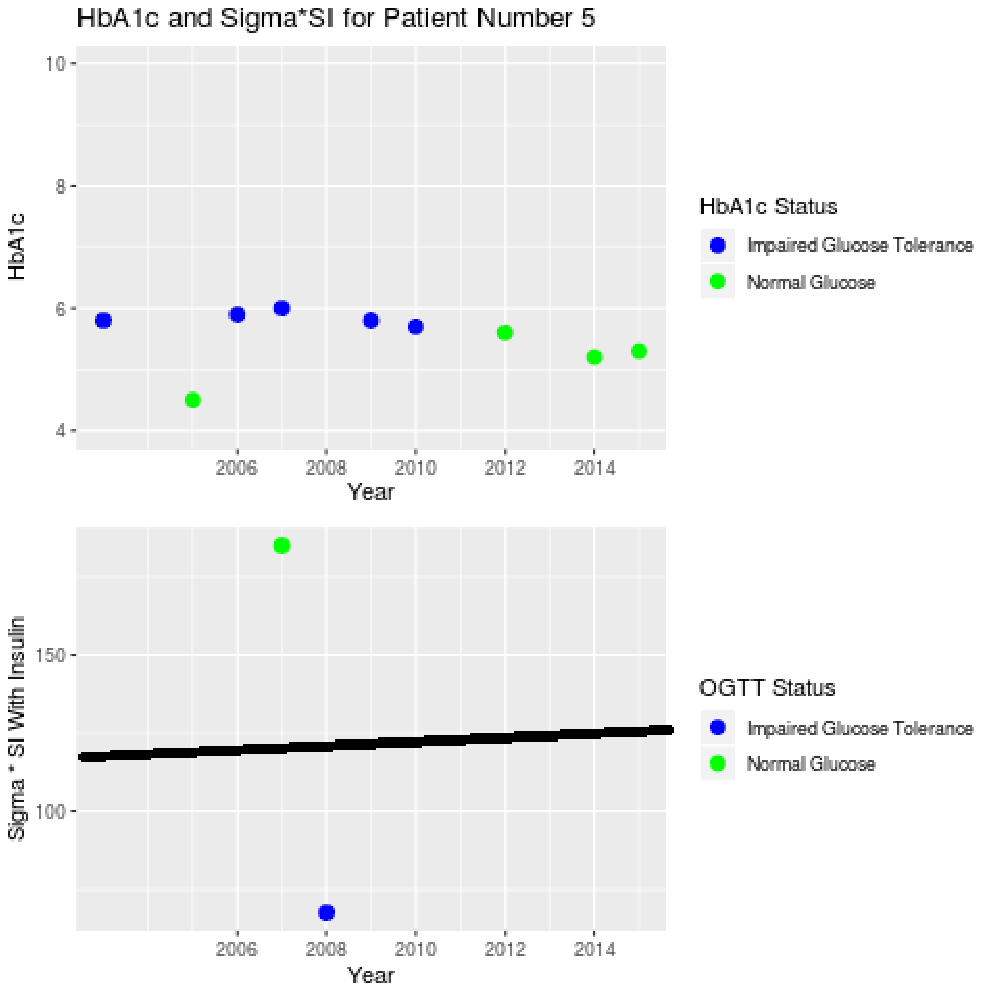}}
\caption{Scatterplot of Parameters and HbA1c by Year}\label{fig:App05}
\end{figure}

\begin{figure}[!tpb]
\centerline{\includegraphics[width = 1\linewidth]{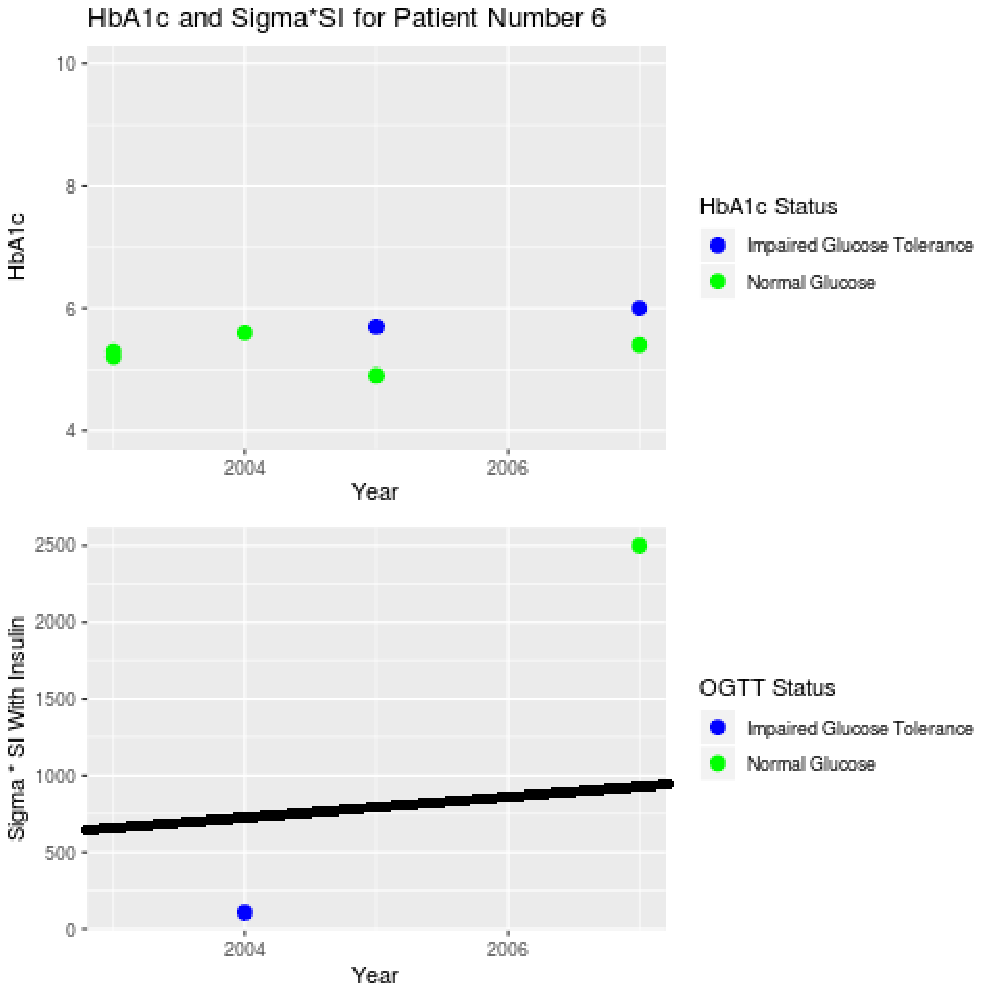}}
\caption{Scatterplot of Parameters and HbA1c by Year}\label{fig:App06}
\end{figure}

\begin{figure}[!tpb]
\centerline{\includegraphics[width = 1\linewidth]{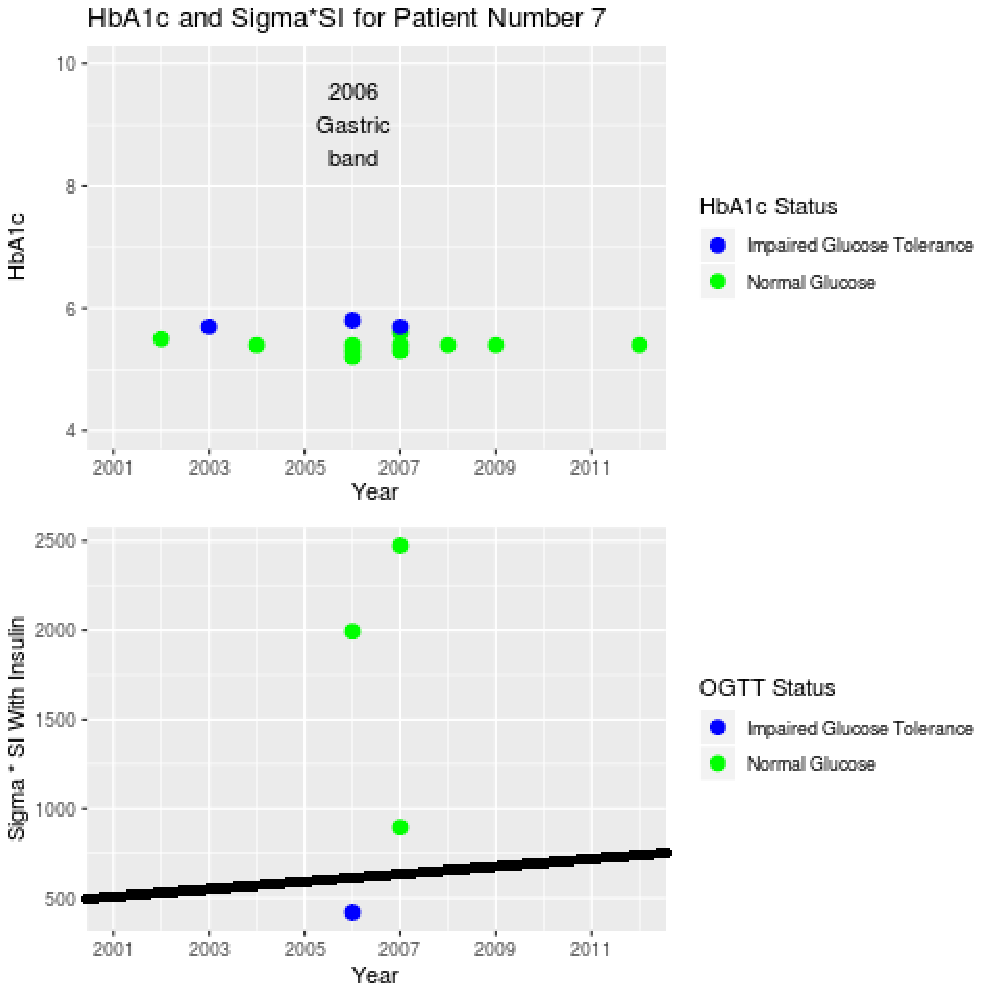}}
\caption{Scatterplot of Parameters and HbA1c by Year}\label{fig:App07}
\end{figure}

\begin{figure}[!tpb]
\centerline{\includegraphics[width = 1\linewidth]{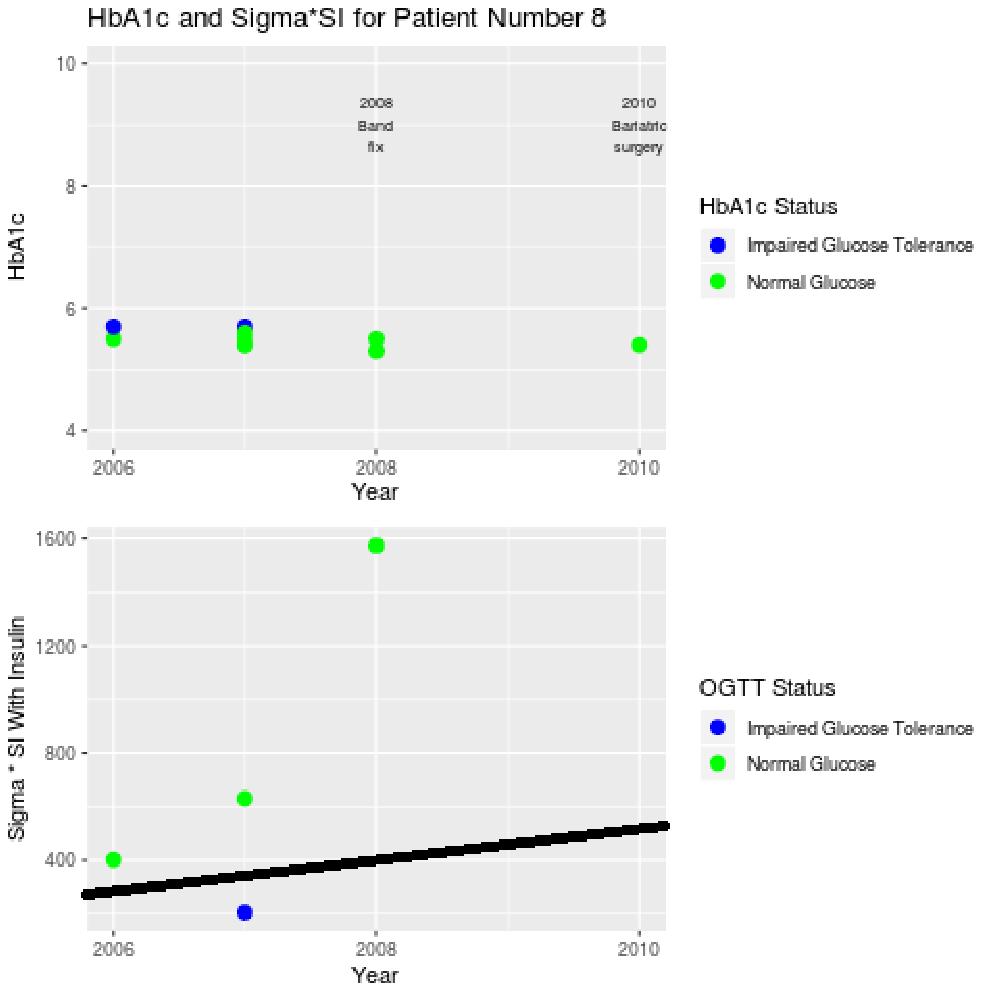}}
\caption{Scatterplot of Parameters and HbA1c by Year}\label{fig:App08}
\end{figure}

\begin{figure}[!tpb]
\centerline{\includegraphics[width = 1\linewidth]{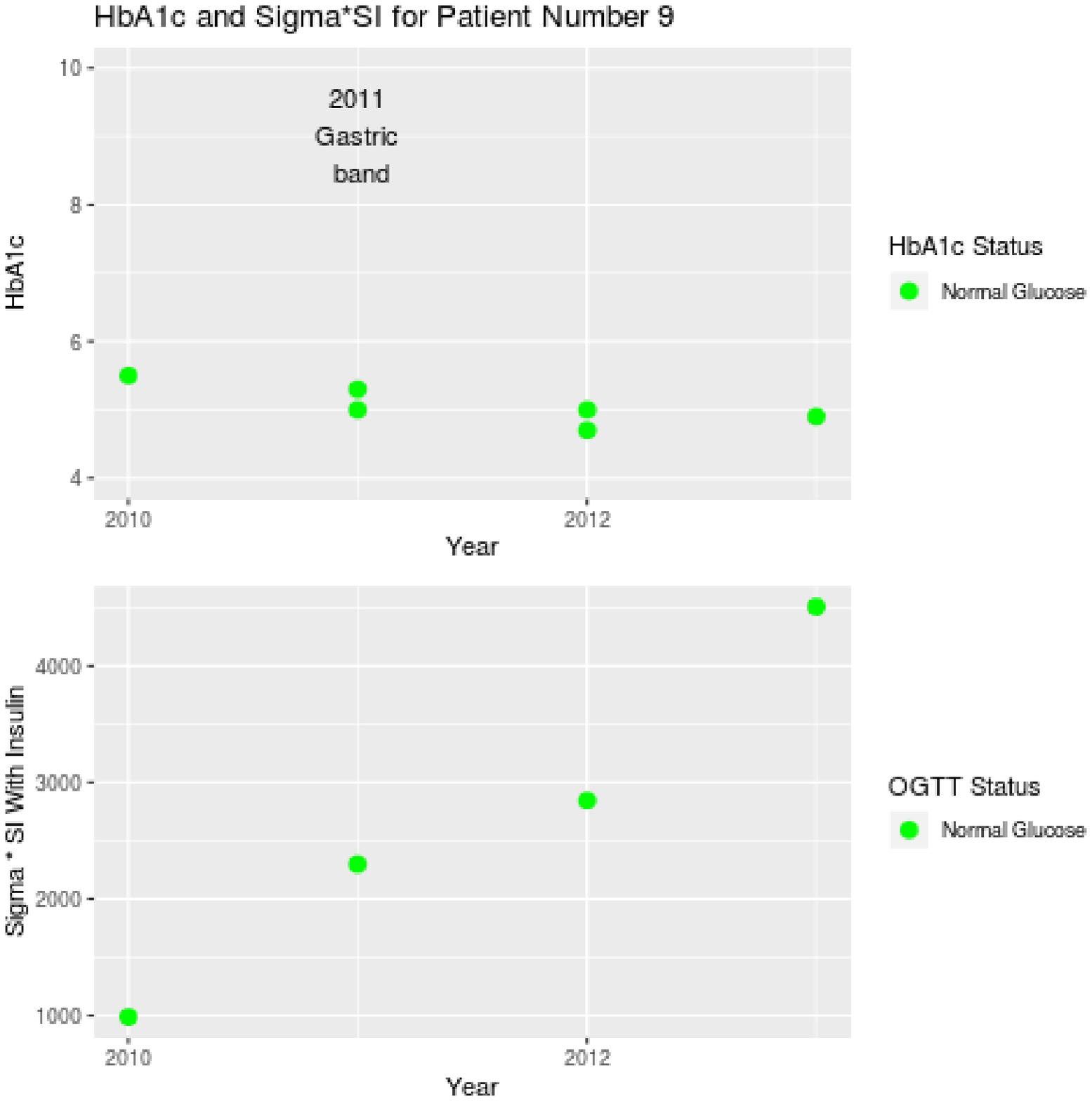}}
\caption{Scatterplot of Parameters and HbA1c by Year}\label{fig:App09}
\end{figure}

\end{document}